\documentstyle[12pt]{article} 
\input psfig.sty
\def\Re{{\cal R \mskip-4mu \lower.1ex \hbox{\it e}\,}}
\def\Im{{\cal I \mskip-5mu \lower.1ex \hbox{\it m}\,}}
\def\nn{\noindent}
\def\ie{{\it i.e.}}
\def\eg{{\it e.g.}}
\def\etc{{\it etc}}
\def\etal{{\it et al.}}

\def\sub#1{_{\lower.25ex\hbox{$\scriptstyle#1$}}}
\def\sul#1{_{\kern-.1em#1}}
\def\sll#1{_{\kern-.2em#1}}  
\def\sbl#1{_{\kern-.1em\lower.25ex\hbox{$\scriptstyle#1$}}}
\def\ssb#1{_{\lower.25ex\hbox{$\scriptscriptstyle#1$}}}
\def\sbb#1{_{\lower.4ex\hbox{$\scriptstyle#1$}}}

\def\to{\rightarrow}
\def\mh{\ifmmode m\sbl H \else $m\sbl H$\fi}
\def\mch{\ifmmode m_{H^\pm} \else $m_{H^\pm}$\fi}
\def\mt{\ifmmode m_t\else $m_t$\fi}
\def\mc{\ifmmode m_c\else $m_c$\fi}
\def\mz{\ifmmode M_Z\else $M_Z$\fi}
\def\mw{\ifmmode M_W\else $M_W$\fi}
\def\mws{\ifmmode M_W^2 \else $M_W^2$\fi}
\def\mhs{\ifmmode m_H^2 \else $m_H^2$\fi}   
\def\mzs{\ifmmode M_Z^2 \else $M_Z^2$\fi}
\def\mts{\ifmmode m_t^2 \else $m_t^2$\fi}
\def\mcs{\ifmmode m_c^2 \else $m_c^2$\fi}
\def\mchs{\ifmmode m_{H^\pm}^2 \else $m_{H^\pm}^2$\fi}
\def\ztwo{\ifmmode Z_2\else $Z_2$\fi}
\def\zone{\ifmmode Z_1\else $Z_1$\fi}
\def\mtwo{\ifmmode M_2\else $M_2$\fi}
\def\mone{\ifmmode M_1\else $M_1$\fi}
\def\tb{\ifmmode \tan\beta \else $\tan\beta$\fi}
\def\xw{\ifmmode x\sub w\else $x\sub w$\fi}
\def\ch{\ifmmode H^\pm \else $H^\pm$\fi}
\def\lum{\ifmmode {\cal L}\else ${\cal L}$\fi}
\def\inpb{\ifmmode {\rm pb}^{-1}\else ${\rm pb}^{-1}$\fi}
\def\infb{\ifmmode {\rm fb}^{-1}\else ${\rm fb}^{-1}$\fi}
\def\epem{\ifmmode e^+e^-\else $e^+e^-$\fi}
\def\ppb{\ifmmode \bar pp\else $\bar pp$\fi}

\def\bsg{\ifmmode b\rightarrow s\gamma \else $b\rightarrow s\gamma$\fi}
 
\newskip\zatskip \zatskip=0pt plus0pt minus0pt
\def\matth{\mathsurround=0pt}

\def\atversim#1#2{\lower0.7ex\vbox{\baselineskip\zatskip\lineskip\zatskip
  \lineskiplimit 0pt\ialign{$\matth#1\hfil##\hfil$\crcr#2\crcr\sim\crcr}}}

\renewcommand{\thefootnote}{\fnsymbol{footnote}}

\begin{document} \begin{titlepage} 
\setcounter{page}{1}
\thispagestyle{empty}
\rightline{\vbox{\halign{&#\hfil\cr
&SLAC-PUB-7367\cr
&December 1996\cr}}}
\vspace{0.8in} 
\begin{center}

{\Large\bf
An Exploration of Below Threshold $Z'$ Mass and Coupling 
Determinations at the NLC}
\footnote{Work supported by the Department of 
Energy, contract DE-AC03-76SF00515.} 
\medskip

\normalsize THOMAS G. RIZZO
\\ \smallskip
{\it {Stanford Linear Accelerator Center\\Stanford University, 
Stanford, CA 94309}}\\ 

\end{center} 

\begin{abstract}
 
We examine of the capability of the Next Linear Collider to determine the 
mass as well as the couplings to leptons and $b$-quarks of a new neutral 
gauge boson, $Z'$, below direct production threshold. By using 
simulated data collected at several different values of $\sqrt s$, we 
demonstrate how this can be done in a model-independent manner via an 
anonymously case approach. The importance of beam polarization to the success 
of this program is discussed. The procedure is shown to be easily extended 
to the case of top and charm quark couplings.

\end{abstract}

\vskip0.45in
\begin{center}

Submitted to Physical Review {\bf D}.

\end{center}


\renewcommand{\thefootnote}{\arabic{footnote}} \end{titlepage}

  
\section{{\bf Introduction}}

While the Standard Model(SM) is in full agreement with all experimental 
data{\cite {warsaw}}, it is generally believed that new physics must exist at 
a scale not far beyond the reach of existing accelerators. Associated with 
this scale may be a host of new and exotic particles. 
A new neutral gauge boson, $Z'$, is the most well-studied of all exotic 
particles and is the hallmark signature for extensions of the SM gauge group. 
Current direct searches for the existence of such particles at the 
Tevatron{\cite {tev}} 
suggest that their masses must be in excess of 500-700 GeV depending upon 
their couplings to the SM fermions and their kinematically accessible 
decay modes. 
If such a particle is found at future colliders the next step will be to 
ascertain its couplings to all of the conventional fermions. In this way, we 
may hope to identify whether this new particle corresponds to any one of the 
many $Z'$'s proposed in the literature or is something else entirely. 
At hadron colliders, 
a rather long list of observables has been proposed over the years 
to probe these 
couplings -- each with its own limitations{\cite {rev1,rev2,rev5}}.  
It has been shown under idealized conditions, 
at least within the context of $E_6$-inspired 
models, that the LHC($\sqrt s=14$ TeV, $100fb^{-1}$) will be able 
to extract useful 
information on all of the $Z'$ couplings for $M_{Z'}$ below $\simeq 1-1.5$ 
TeV. 
It is {\it not} clear, however, how much of this program can be carried out 
using realistic detectors at the LHC{\cite {rev5}} and how well it generalizes  
to other extended gauge models since detailed simulation studies have yet to be 
performed. 

At the NLC, when $\sqrt s < M_{Z'}$ (the most likely scenario for a first 
generation machine given the 
Tevatron bounds) a $Z'$ can only manifest itself 
indirectly as deviations in, \eg, cross sections and asymmetries from their SM 
expectations. This is analogous to the observation of the SM $Z$ at 
PEP/PETRA/TRISTAN energies through deviations from the expectations of QED. 
Fortunately the list 
of useful precision measurements that can be performed at the NLC is 
reasonably long and the expected large beam polarization($P$)  
plays an important role--essentially doubling the number of useful observables. 
In the past, analyses of the ability of 
the NLC to extract $Z'$ coupling information in this situation have taken 
for granted that the 
value of $M_{Z'}$ is already known from elsewhere, \eg, the 
LHC{\cite {rev3,rev4}}. (In fact, one might argue that if a 1 TeV $Z'$ is 
discovered at the LHC, a future lepton linear collider designed to sit on 
this $Z'$ must 
be built and will thus quite easily determine all of the $Z'$ couplings 
in analogy to SLC and LEP.) 
Here we address the more complex issue of whether it is possible for the NLC to 
obtain information on couplings of the $Z'$ if the mass were for some 
reason {\it a priori} unknown. In this case we would not only want to 
determine couplings but the $Z'$ mass as well. We will limit our discussion 
below to the $e^+e^-$ channel and ignore the additional information available 
through $e^-e^-$ collisions{\cite {cuy}}.

If the $Z'$ mass were unknown it would appear that the traditional NLC $Z'$ 
coupling analyses would become problematic. Given a set of 
data at a fixed value of $\sqrt s$ which shows deviations from the SM, one 
would not be able to {\it simultaneously} extract the value of $M_{Z'}$ as 
well as the corresponding couplings. The reason is clear: to leading order in 
$s/M_{Z'}^2$, rescaling 
all of the couplings and the value of $M_{Z'}$ by an overall common factor 
would leave the observed deviations from the SM invariant. In this 
approximation, the $Z'$ exchange appears only as a simple contact interaction. 
Thus as long as $\sqrt s < M_{Z'}$, the only potential solution to this 
problem lies in obtaining data on deviations from the 
SM at {\it several}, distinct $\sqrt s$ values and combining them into a 
single fit. It is clear from the beginning that all of the set of 
$\sqrt s$ values 
chosen for this analysis cannot lie too far below the $Z'$ mass otherwise we 
would always remain in the contact interaction limit. It must be, for at 
least one of the $\sqrt s$ choices, that sub-leading terms of relative order 
$s/M_{Z'}^2$ are of numerical importance. This suggests that the maximum value 
of $\sqrt s$ within this set should only be about a factor of 2-3 lower than 
the $Z'$ mass.

Here we report on the first analysis 
of this kind, focussing on observables involving only leptons and/or  
$b$-quarks. In performing such an analysis, we need to know how many $\sqrt s$ 
values are needed. We need to know how we distribute the integrated 
luminosity($\cal L$) to optimize the results. Similarly, we must address 
whether such an analysis can be performed while maintaining model-independence. 
In this {\it initial} study we begin to address these and some related 
issues. It is clear that that there is a lot more to be done before we 
have the answers to all these questions.

\section{{\bf Analysis}}

In order to proceed with this benchmark study, we will make a number of 
simplifying assumptions and parameter choices. These can be modified at a 
later stage to see how they influence our results. (The basic analysis 
follows that discussed in {\cite {hr,physrep}}.) 
In this analysis we consider the following 
ten observables: the total production cross sections for leptons and 
$b$-quarks, $\sigma_{\ell,b}$, the corresponding forward-backward 
asymmetries, $A_{FB}^{\ell,b}$, the left-right asymmetry obtained from 
flipping the initial electron beam polarization, $A_{LR}^{\ell,b}$, and the 
polarized forward-backward asymmetry, 
$A_{pol}^{FB}(\ell,b)$. For $\tau^+\tau^-$ final states, we include the 
average $\tau$ polarization, $<P_\tau>$, as well as the forward-backward 
asymmetry in the $\tau$ polarization, $P_\tau^{FB}$. Other inputs and 
assumptions are summarized as follows:

\begin{tabbing}
   Anomalous Trilinear Gauge Couplings Fun \= 8.Fun and games with Extended 
Technicolor models \kill
   ~~~~~~~e,$\mu$,$\tau$ universality \> ISR with $\sqrt {s'}/\sqrt {s} >0.7$\\
   ~~~~~~~$P=90\%$, $\delta P/P=0.3\%$  \> $\delta {\cal L}/ {\cal L}=0.25\%$\\
   ~~~~~~~$\epsilon_b=50\%$, $\Pi_b=100\%$  \>   $|\theta|>10^\circ$\\
   ~~~~~~~$\epsilon_{e,\mu,\tau}(\sigma)=100\%$, $\epsilon_\tau(P_\tau)=50\%$ 
   \> Neglect $t$-channel exchange in $e^+e^-\to e^+e^-$ 
\end{tabbing}

\noindent Of special note on this list of assumptions are: ($i$) a $b$-tagging 
efficiency($\epsilon_b$) of $50\%$ for a purity($\Pi_b$) of 100$\%$, ($ii$) 
the efficiency for identifying all leptons is assumed to be 100$\%$, although 
only $50\%$ of $\tau$ decays are assumed to be polarization analyzed, ($iii$) 
a $10^\circ$ angle cut has been applied to all final state fermions to mimic 
the anticipated detector acceptance, 
($iv$) a strong energy cut to remove events with an excess of initial state 
radiation(ISR) has been made--this is critical since events with lower 
effective values of $\sqrt s$ substantially dilute our sensitivity, 
($v$) it has been assumed that both the 
beam polarization($P$) and machine luminosity($\cal L$) are both well 
measured. It is important to note that we have 
{\it not} included the $t$-channel contributions to $e^+e^- \to e^+e^-$ in 
these calculations. 
In addition to the above, final state QED as well as QCD corrections are 
included, the 
$b$-quark and $\tau$ masses have been neglected, and the possibility of any 
sizeable $Z-Z'$ mixing has also been neglected; this is an 
excellent approximation for the $Z'$ mass range of interest to us given that 
we are not interested in the $Z'\to W^+W^-$ mode. Since 
our results will generally be statistics limited, the role played by the 
systematic uncertainties associated with the parameter choices above will 
generally be rather minimal, especially in the lepton case. Larger systematics 
should possibly be associated with the $b$-quark final states{\cite {dis}} 
but they have been ignored here for simplicity. 

To insure model-independence, the values of the $Z'$ couplings, \ie, 
$(v,a)_{\ell,b}$, as well as $M_{Z'}$, are chosen {\it randomly} and 
{\it anonymously} using a random number generator from 
rather large ranges representative of a number of extended gauge models. 
Monte Carlo data representing the above observables 
is then generated for several different values of $\sqrt s$. At this point, the 
values of the mass and couplings are not `known'  
{\it a priori}, but will later be compared with what is extracted 
from the Monte Carlo generated event sample. Following this approach 
there is no particular relationship between any of the couplings and 
there is no dependence upon or relationship to any particular 
$Z'$ model. (We chose to normalize our couplings so that 
for the SM $Z$, $a_\ell=-1/2$.) Performing this analysis for a wide range of 
possible mass and coupling choices then shows the power as well as the 
potential limitations of this technique.

To get an understanding for how this procedure works in general we will make 
three representative case studies for the $Z'$ mass and couplings, labelled 
here by I, II and III. 
There is nothing special about these three choices and several other parameter 
sets have been analyzed in comparable detail to show that the results that 
we display below are rather typical. To begin our analysis, let us try 
choosing three distinct $\sqrt s$ values. Specifically, we 
generate Monte Carlo `data' at $\sqrt {s}=$0.5, 0.75 and 1 TeV 
with associated integrated luminosities of 70, 100, and 150 
$fb^{-1}$, respectively. These luminosities are only slightly larger than 
the typical one year values as conventionally quoted{\cite {zdr}} and assumes 
a reasonable time evolution of the collider's center of mass energy. 
Subsequently, we   
determine the 5-dimensional $95\%$ CL allowed region for the mass and 
couplings from a simultaneous fit to all of the leptonic and $b$-quark 
observables following the input assumptions listed above. This 
5-dimensional region is then projected into a series of 2-dimensional plots 
which we now examine in detail.

\vspace*{-0.5cm}
\nn
\begin{figure}[htbp]
\centerline{
\psfig{figure=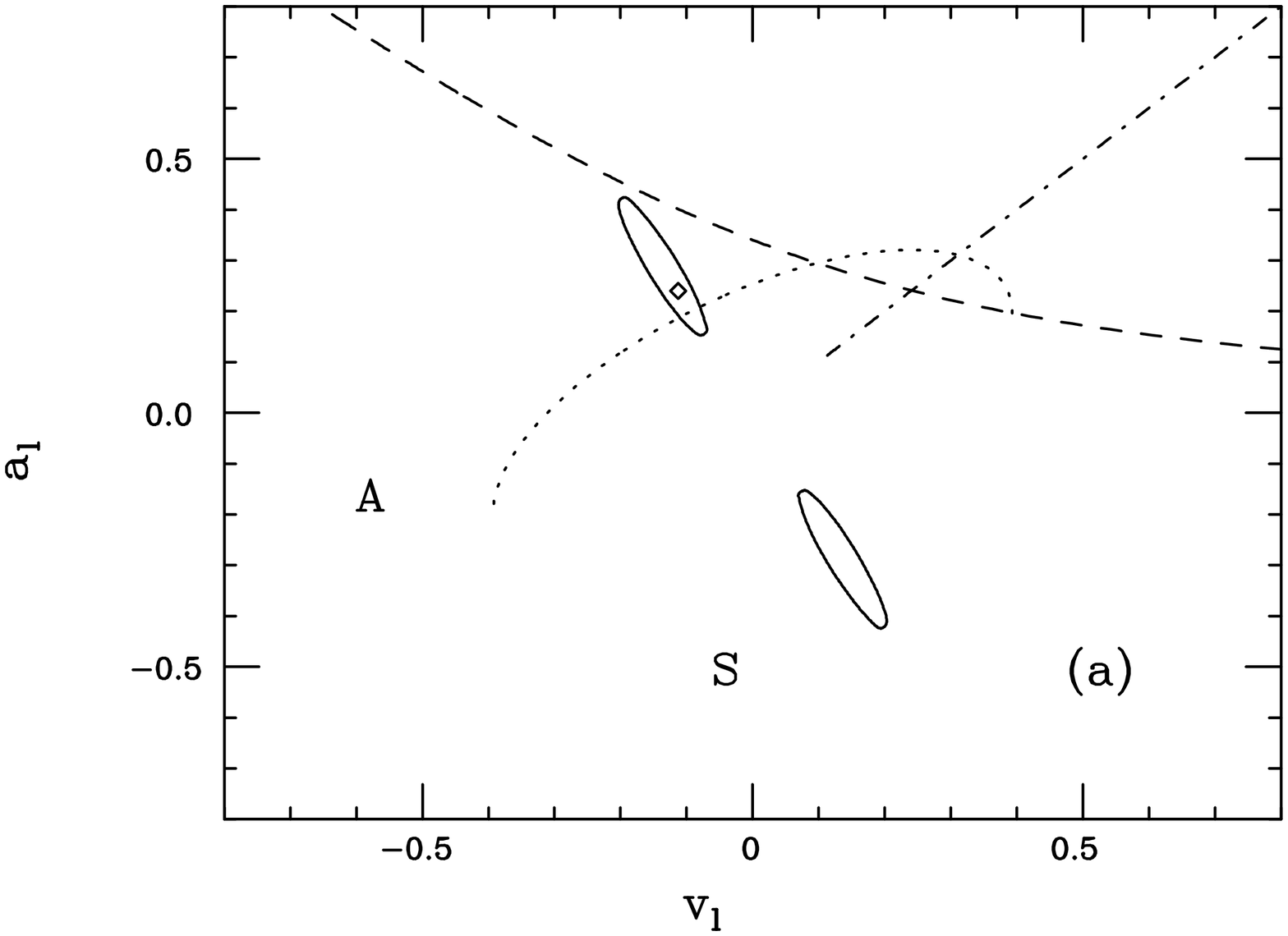,height=9.1cm,width=9.1cm,angle=-90}
\hspace*{-5mm}
\psfig{figure=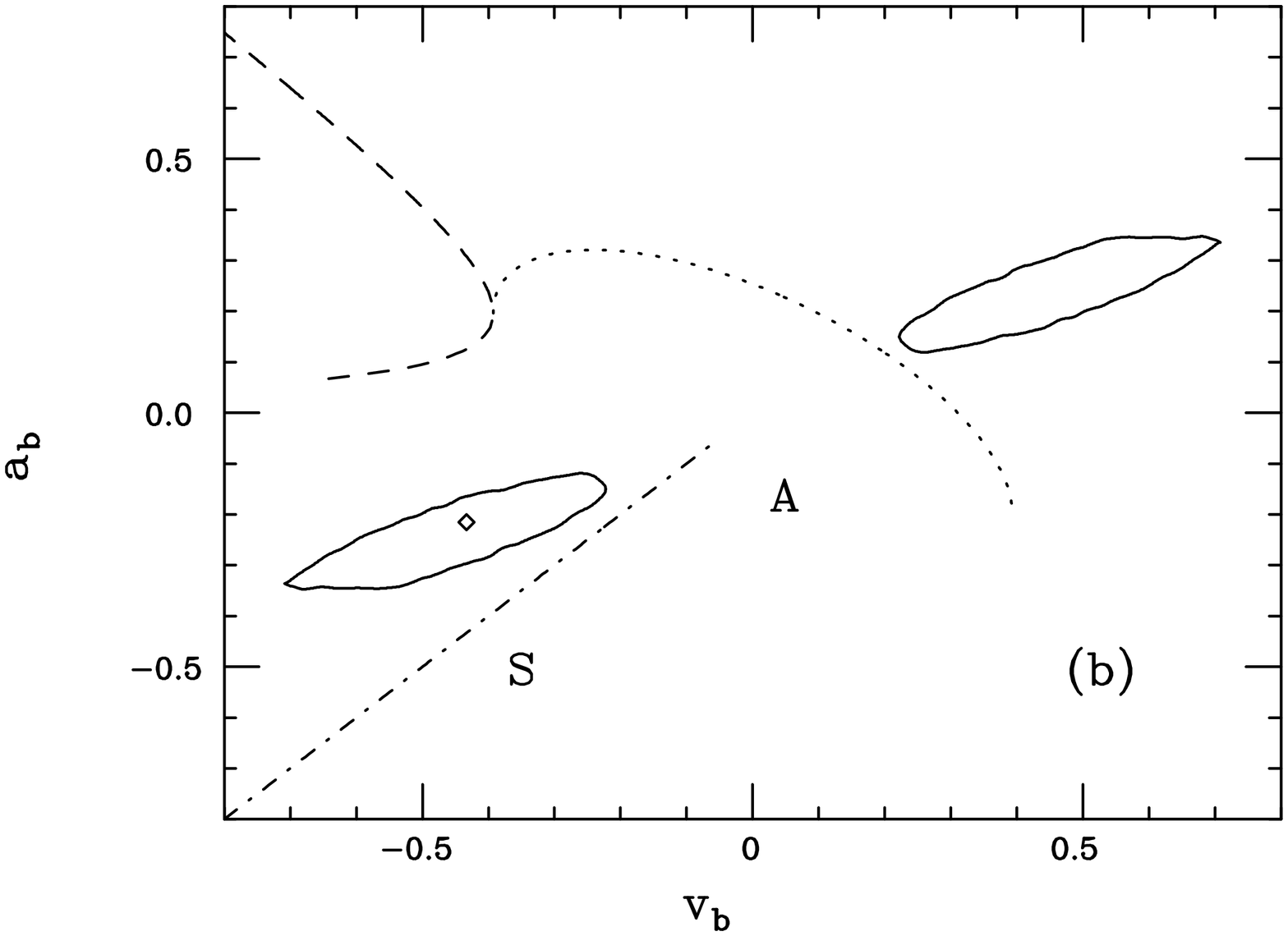,height=9.1cm,width=9.1cm,angle=-90}}
\vspace*{-0.75cm}
\centerline{
\psfig{figure=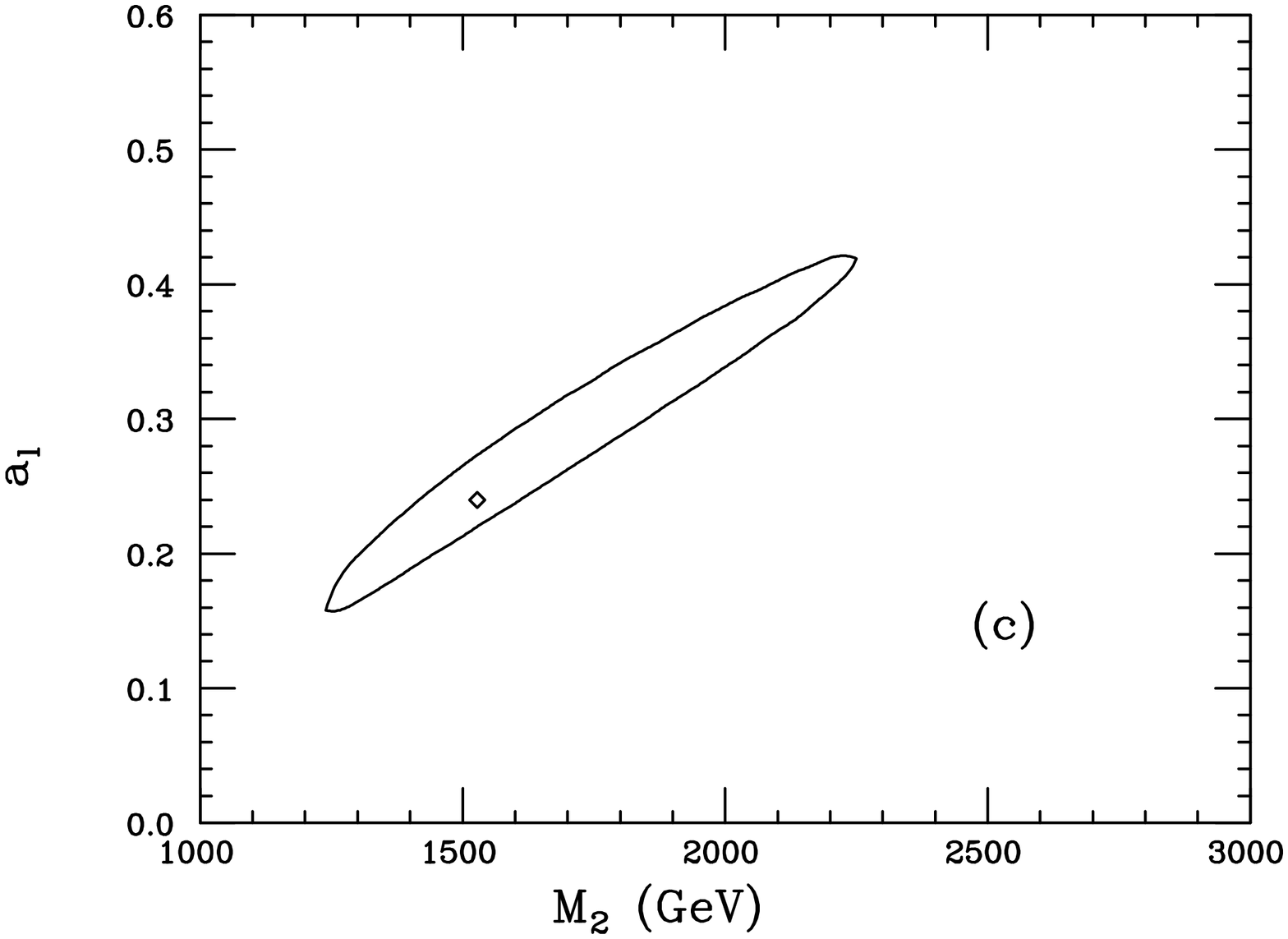,height=9.1cm,width=9.1cm,angle=-90}}
\vspace*{-1cm}
\caption{\small $95\%$ CL allowed regions for the extracted values of the 
(a) lepton and (b) $b$-quark couplings 
for the $Z'$ of case I compared with the predictions of the $E_6$ 
model(dotted), the Left-Right Model(dashed), and the Un-unified 
Model(dash-dot), 
as well as the Sequential SM and Alternative LR Models(labeled by `S' and `A', 
respectively.) (c) Extracted $Z'$ mass; only the $a_\ell >0$ branch is shown. 
In all cases the diamond represents the corresponding input values.}
\end{figure}
\vspace*{-0.5cm}
\nn
\begin{figure}[htbp]
\centerline{
\psfig{figure=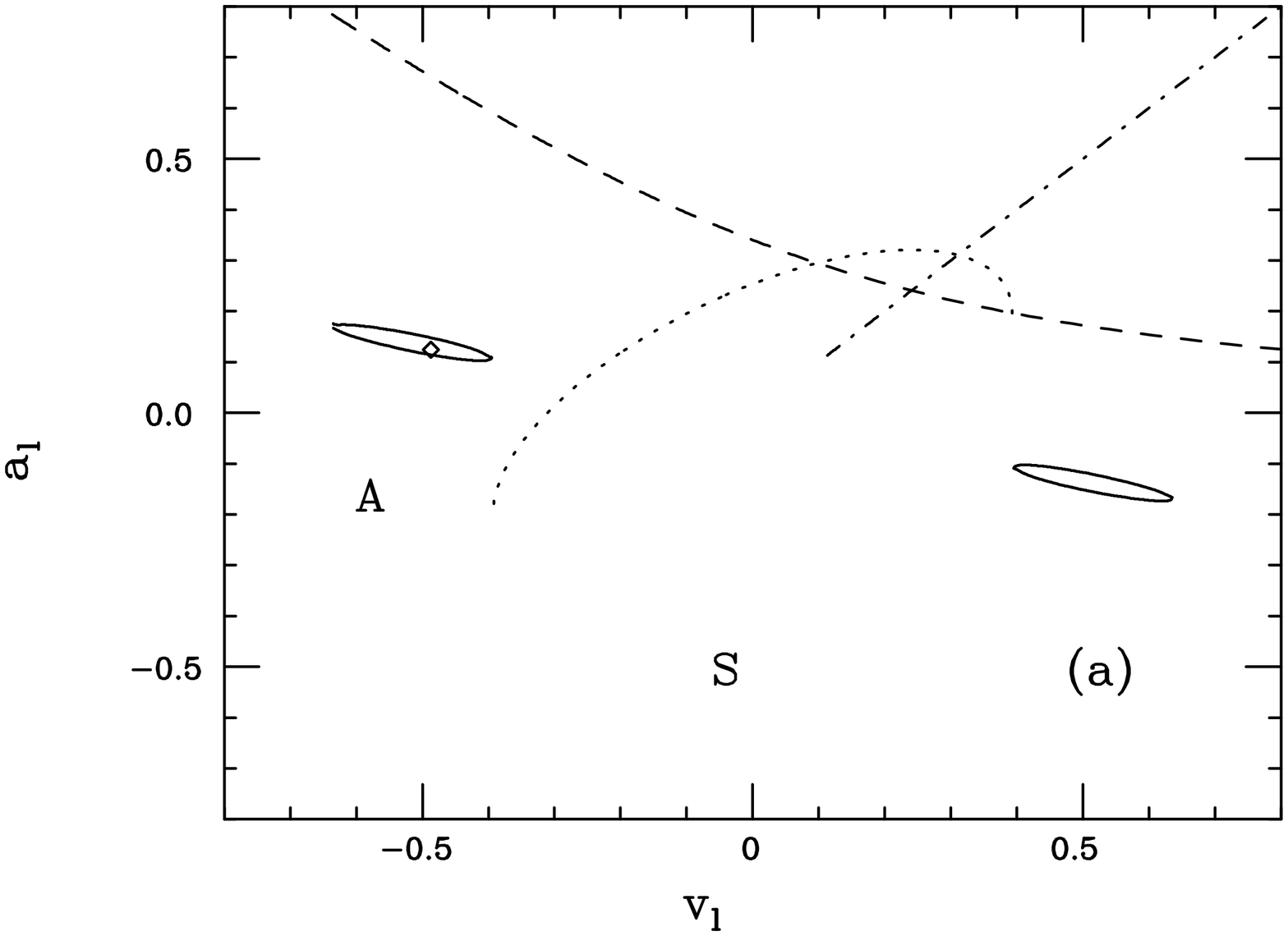,height=9.1cm,width=9.1cm,angle=-90}
\hspace*{-5mm}
\psfig{figure=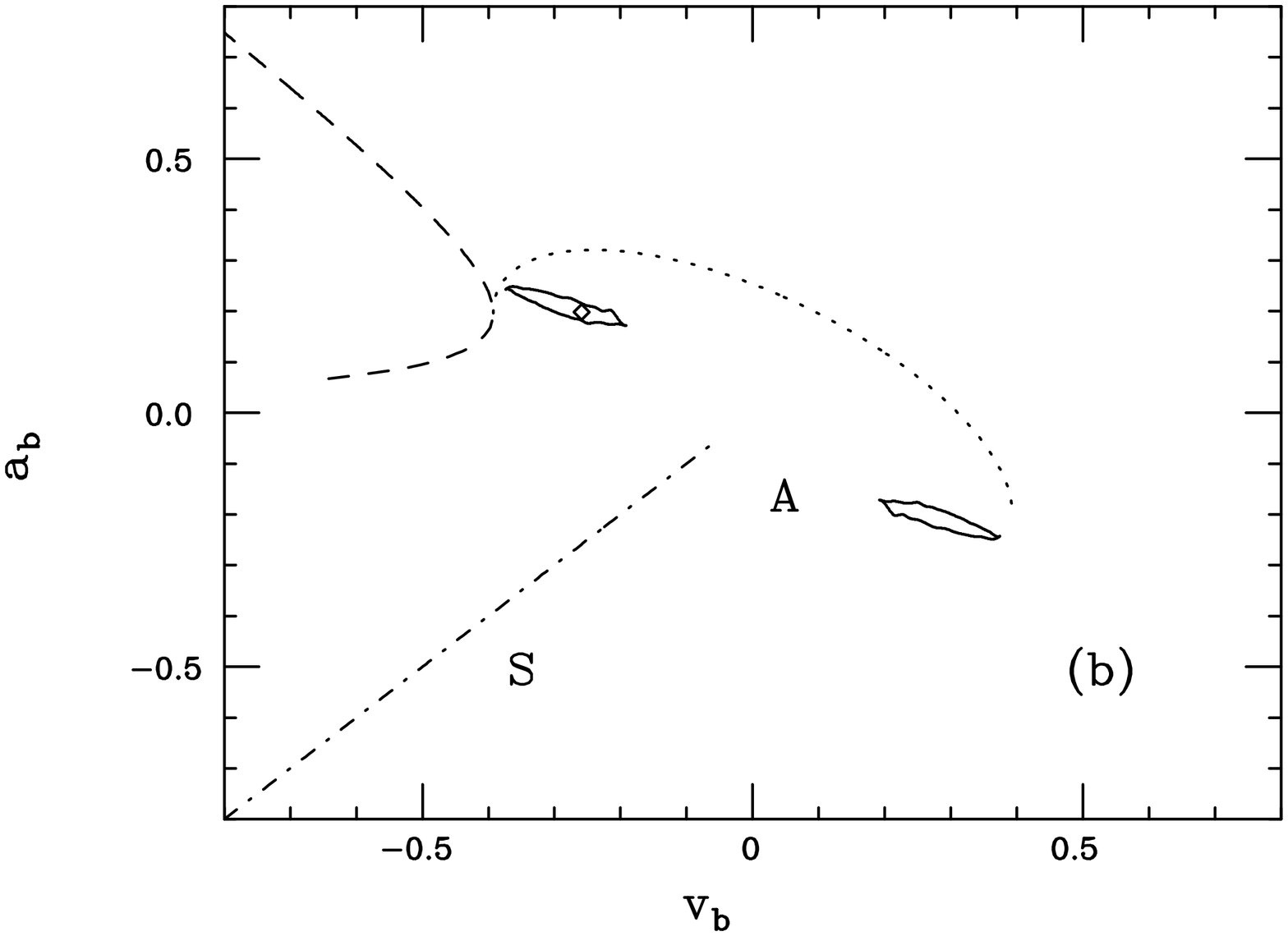,height=9.1cm,width=9.1cm,angle=-90}}
\vspace*{-0.75cm}
\centerline{
\psfig{figure=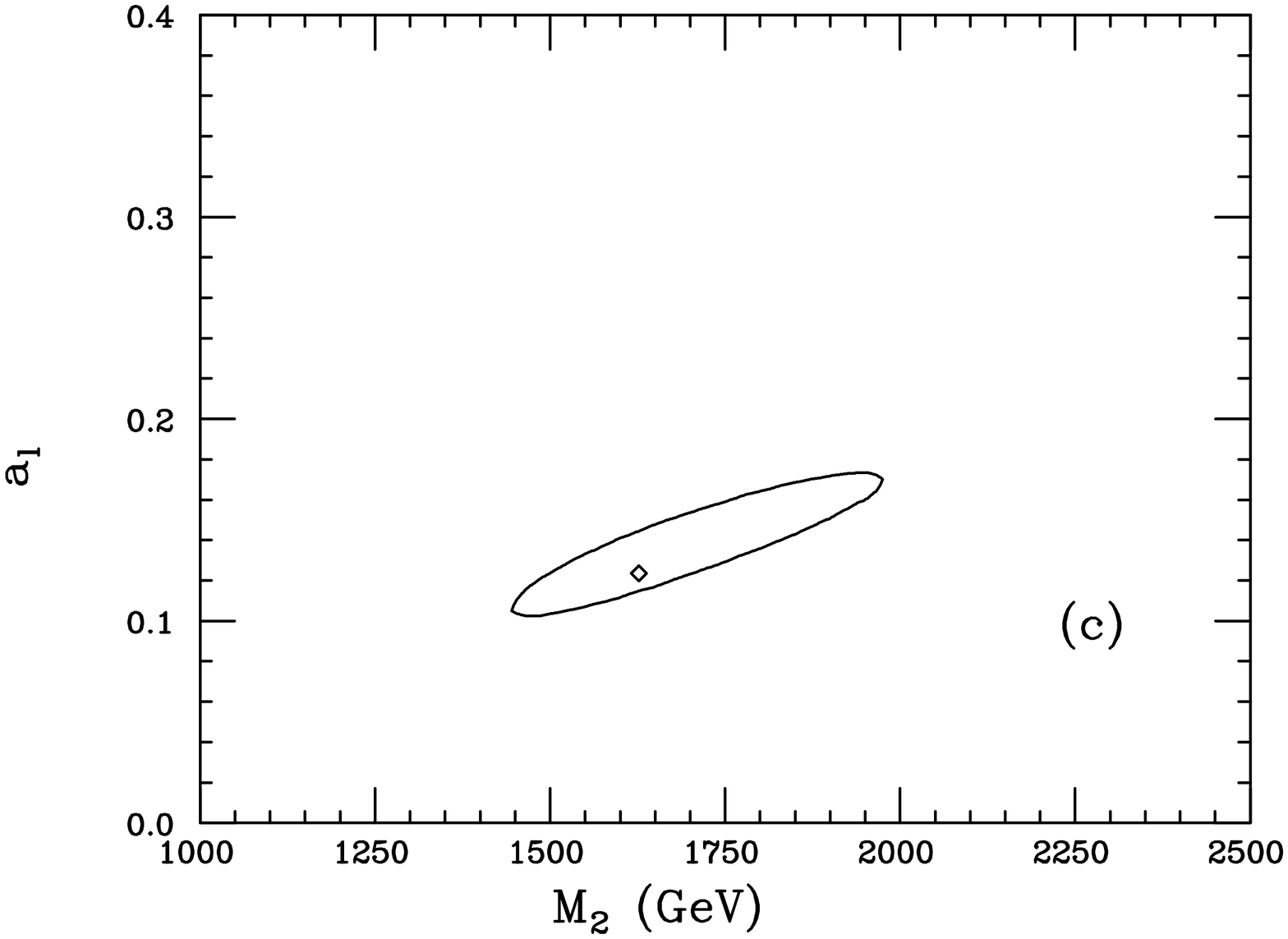,height=9.1cm,width=9.1cm,angle=-90}}
\vspace*{-1cm}
\caption{\small Same as Fig. 1 but for a different choice of $Z'$ mass and 
couplings referred to as case II in the text.}
\end{figure}
\vspace*{-0.5cm}
\nn
\begin{figure}[htbp]
\centerline{
\psfig{figure=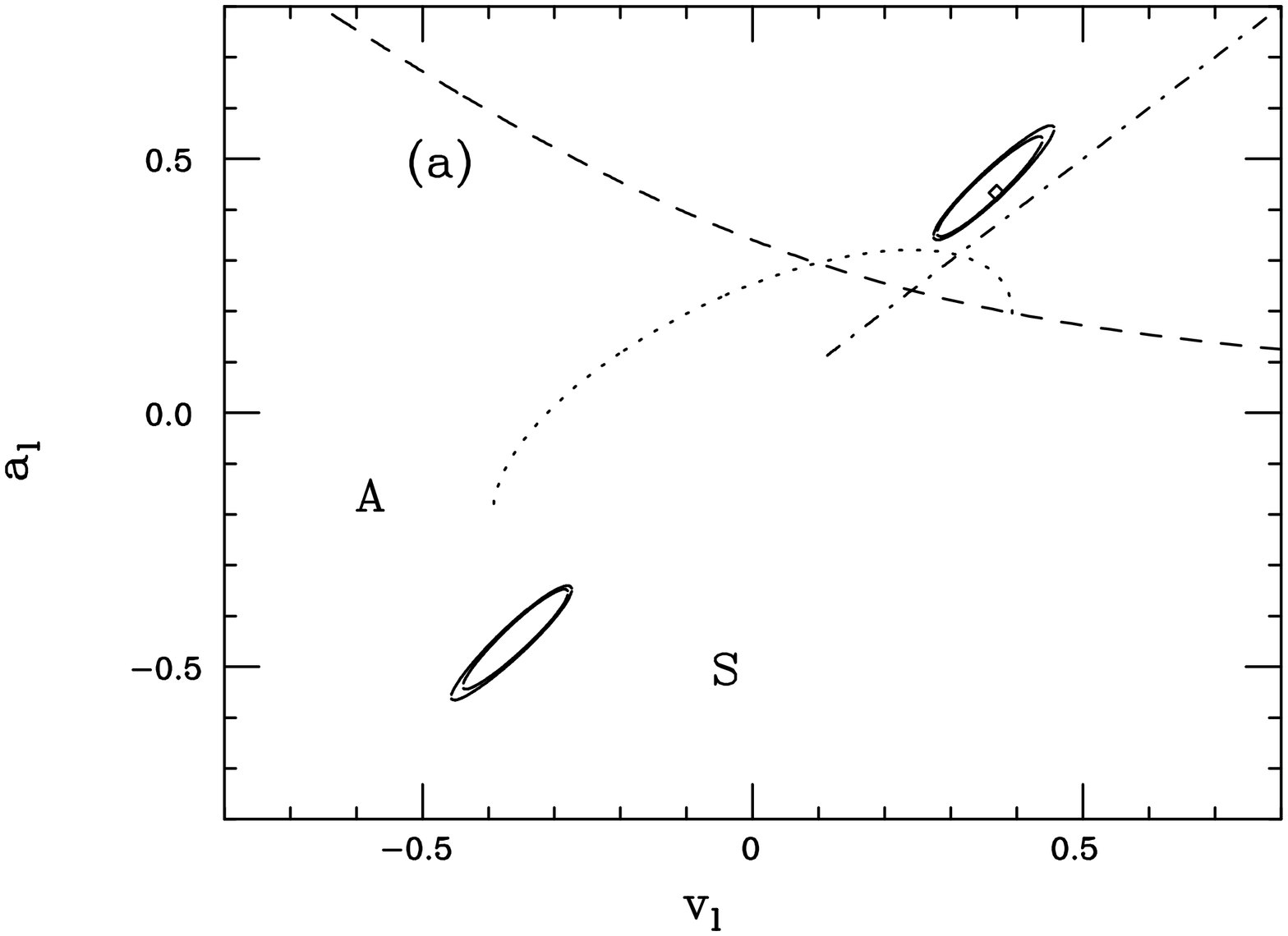,height=9.1cm,width=9.1cm,angle=-90}
\hspace*{-5mm}
\psfig{figure=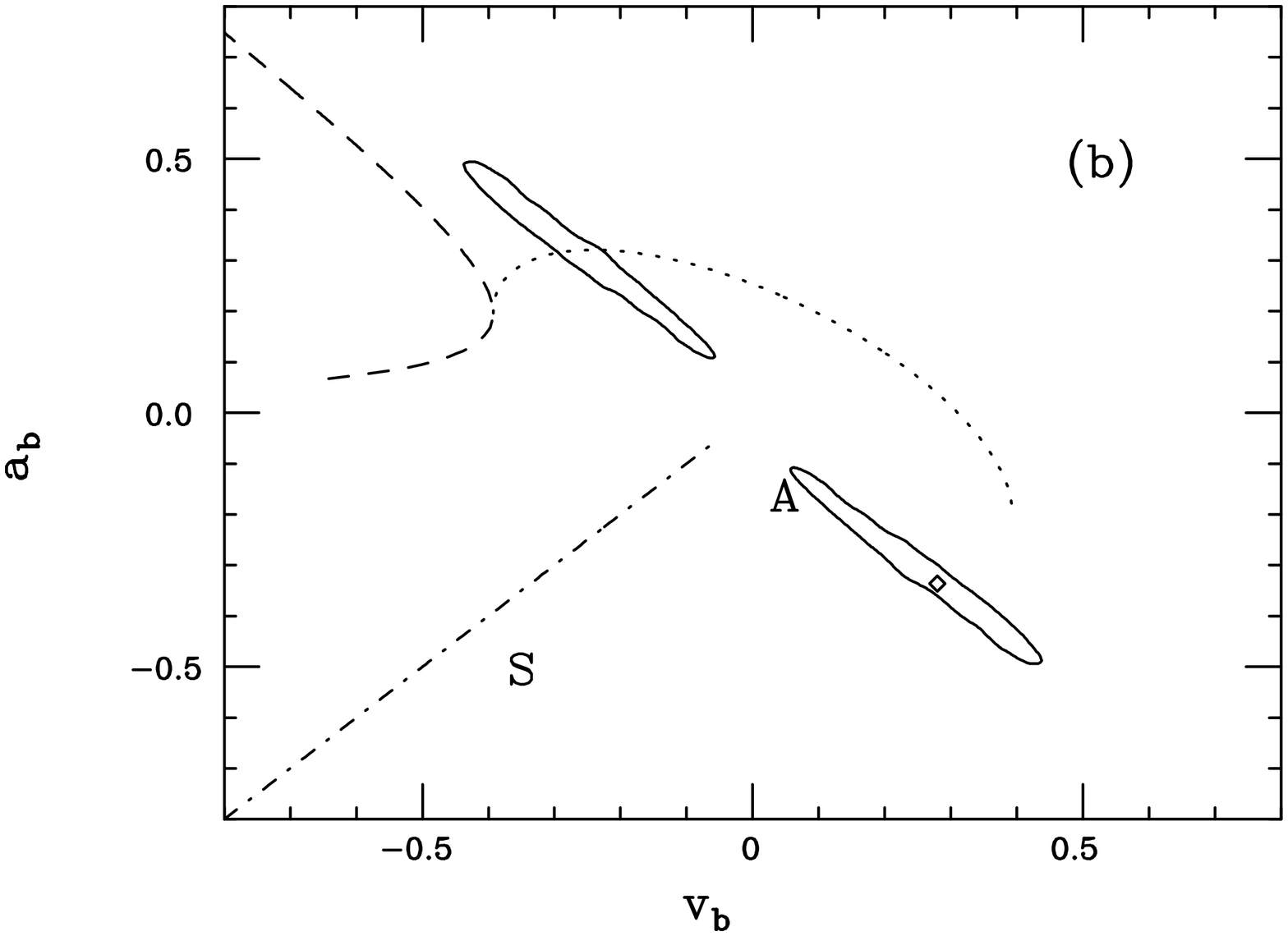,height=9.1cm,width=9.1cm,angle=-90}}
\vspace*{-0.75cm}
\centerline{
\psfig{figure=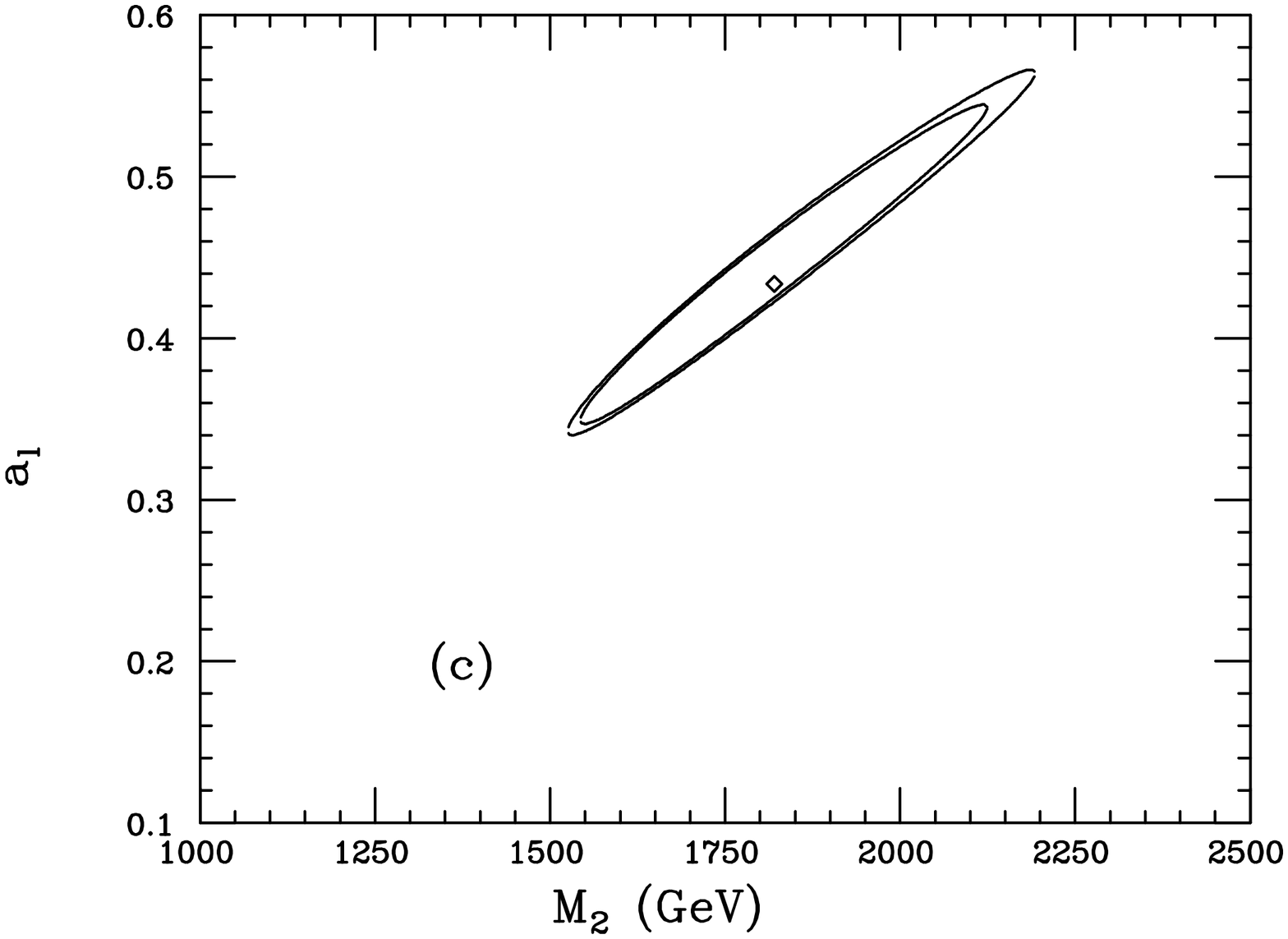,height=9.1cm,width=9.1cm,angle=-90}}
\vspace*{-1cm}
\caption{\small Same as Fig. 1 but for a third choice of $Z'$ mass and 
couplings referred to as case III in the text. The larger(smaller) allowed 
region in each case corresponds to $P=80(90)\%$.}
\end{figure}

Figs. 1-3 show the results of our analysis for these three case studies 
compared 
with the expectations of a number of well-known $Z'$ models{\cite {rev1}}. 
To be specific we have 
considered ({\it i}) the $E_6$ effective rank-5 model(ER5M), which predicts a 
$Z'$ whose couplings depend on a single parameter 
$-\pi/2 \leq \theta \leq \pi/2$; 
({\it ii}) the Sequential Standard Model(SSM) 
wherein the new $W'$ and $Z'$ are just heavy versions of the SM particles (of 
course, this is not a true model in the strict sense but is commonly used as a 
guide by experimenters); ({\it iii}) the Un-unified Model(UUM), based on the 
group $SU(2)_\ell \times SU(2)_q \times U(1)_Y$, which has a 
single free parameter $0.24 \leq s_\phi \leq 0.99$; 
({\it iv}) the Left-Right Symmetric Model(LRM), based on the group 
$SU(2)_L \times SU(2)_R \times U(1)_{B-L}$, 
which also has a free parameter ($\kappa=g_R/g_L\geq 0.55$) of order unity 
which is just the ratio of the gauge couplings 
and, lastly, ({\it v}) the Alternative Left-Right Model(ALRM), based on the 
same extended group as the LRM but now arising from 
$E_6$, wherein the fermion assignments are modified in comparison to the LRM 
due to an ambiguity in how they are embedded in the {\bf 27} representation.

By examining these figures, several things are immediately 
apparent -- the most obvious being that two 
distinct allowed regions are obtained from the fit in all three cases. 
This ambiguity is two-fold and {\it not} four-fold in that there is a unique 
choice of $b$-couplings for a fixed choice of leptonic couplings. 
(Of course as we might hope, the input 
values are seen to lie nicely inside one of them.) This two-fold ambiguity 
occurs due to our inability to make an absolute   
determination of the overall sign of {\it one} of the $Z'$ couplings, \eg, 
$a_\ell$. If the sign 
of  $a_\ell$ were known, only a single allowed region would appear in 
Figs.~1-3 for both leptons and $b$-quarks and a unique coupling 
determination would thus be obtained. 
Note that this {\it same} sign ambiguity arises in SLC/LEP data for the 
SM $Z$ and is only removed through the examination of low-energy neutrino 
scattering. Secondly, we see that the leptonic couplings 
are always somewhat better determined than are those of the $b$-quark, which is 
due to the fact that the leptonic 
observables involve only leptonic couplings, while 
those for $b$-quarks involve both 
types. In addition, there is more statistical power available in the lepton 
channels due to the assumption of universality and the fact that the 
leptonic results employ two additional observables related to the final 
state $\tau$ 
polarization. If the $b$-quark observables had significant systematic errors, 
these allowed regions would become larger still. 
Thirdly, we see in particular from Figs. 1a-b and 3a-b the importance 
in obtaining coupling information for a number of 
different fermion species. If only the Fig. 1a(3b) results were available, one 
might draw the hasty conclusion that an $E_6$-type $Z'$ had been found. 
Fig. 1b(3a)  
clearly shows us that this is not the case. Evidently none of the $Z'$'s 
associated with cases I-III correspond to {\it any} well-known 
model. Fourthly, we note that changing the beam 
polarization from $90\%$ to $80\%$ does not appreciably alter our results. 
Lastly, as promised, the $Z'$ mass is 
determined in all three cases, although with somewhat smaller uncertainties in 
case II. It is important for the reader to realize that there is nothing 
special about any of these three particular cases. 
It is clear from this set of random choices for masses and couplings 
that this procedure should be viable for $Z'$ masses up to about 2 TeV for 
the set of integrated luminosities that we have chosen unless both of the 
leptonic $Z'$ couplings are accidentally small resulting in a reduced 
sensitivity to the existence of the $Z'$. 

It is straightforward to extend this analysis to fermion final states other 
than leptons and the $b$-quark. The extrapolation to charm is the most obvious 
in that apart from identification efficiencies and potentially larger 
systematic errors there is little difference in performing the 
five-dimensional fit with either $b$ or $c$ since the fermion couplings were 
randomly chosen. (Of course we might imagine, however, now doing a more 
ambitious {\it seven}-dimensional fit with all of the couplings being allowed 
to float.) The extension to $t$-quarks would be less straightforward due to 
their large mass and rapid decay to $bW$. In principle, however, the same set 
of observables could be constructed for top as was used in the $b$-quark 
analysis above.

\vspace*{-0.5cm}
\nn
\begin{figure}[htbp]
\centerline{
\psfig{figure=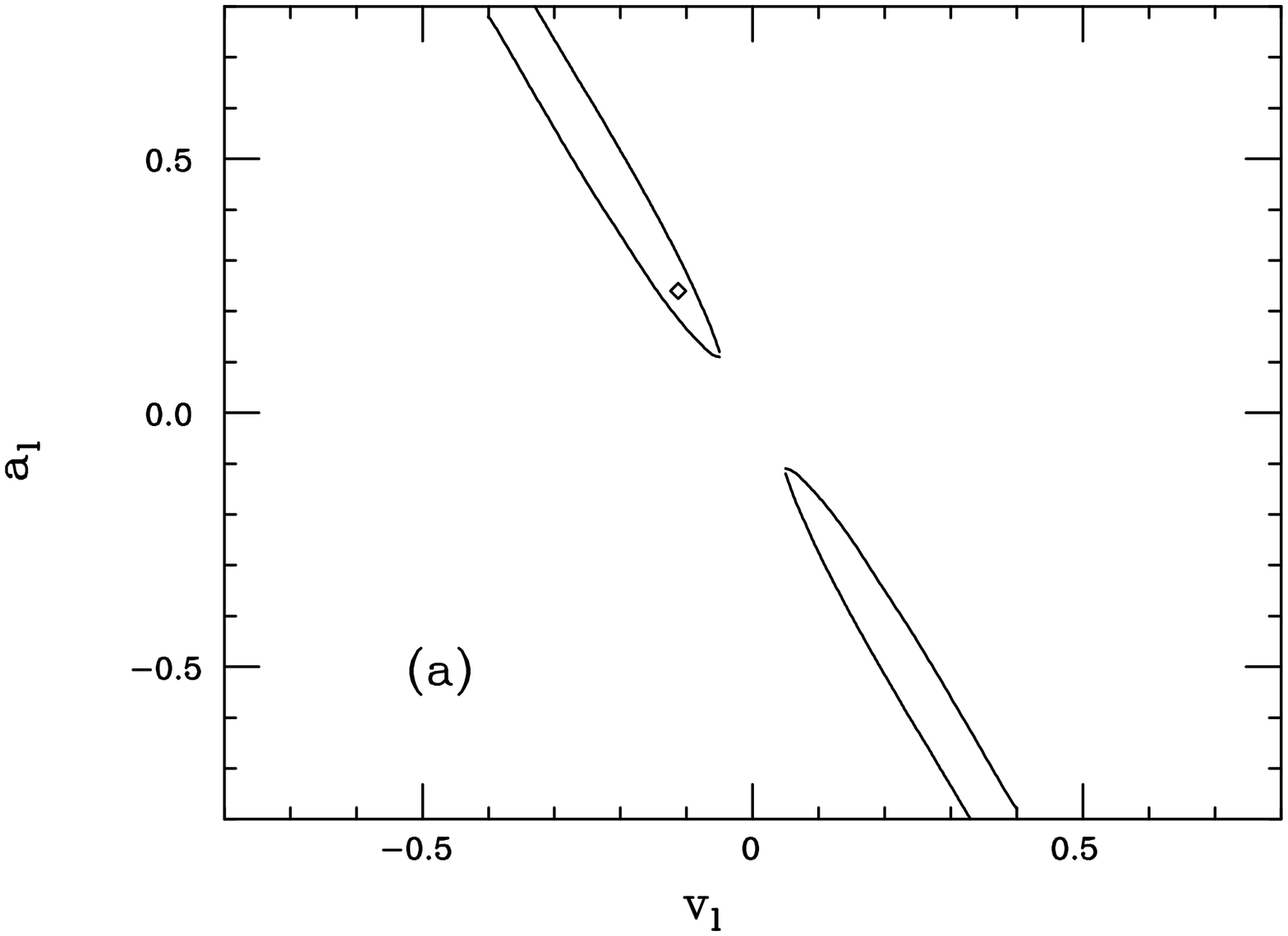,height=9.1cm,width=9.1cm,angle=-90}
\hspace*{-5mm}
\psfig{figure=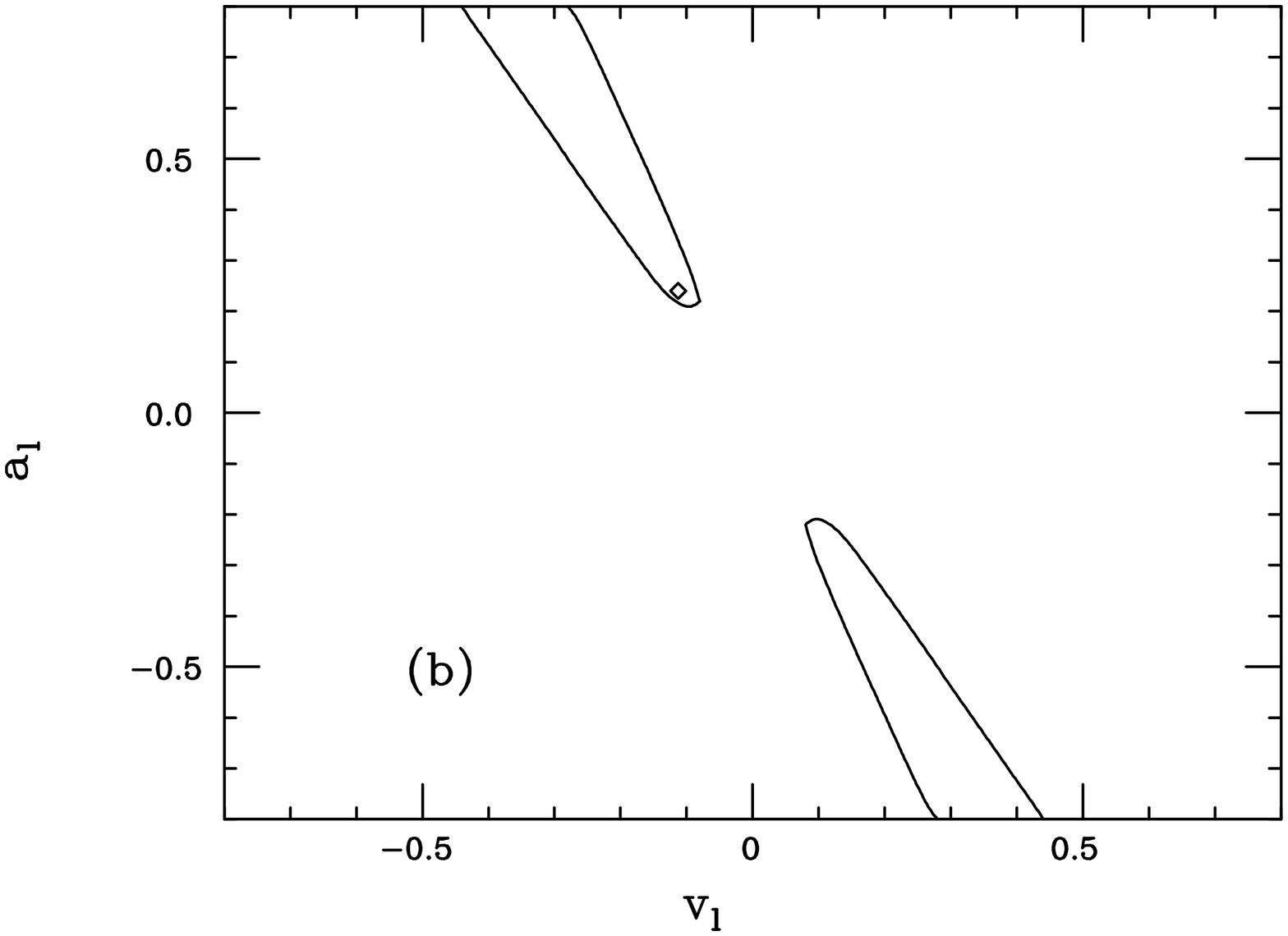,height=9.1cm,width=9.1cm,angle=-90}}
\vspace*{-1cm}
\caption{\small Failure of the method in case I when data is taken at 
(a) too few (`2-point' fit) or (b) too many (`6-point' fit) different 
center of mass energies for the same total integrated 
luminosity as in Figs. 1-3. The luminosities are distributed as discussed 
in the text.}
\end{figure}
\vspace*{0.4mm}

Of course, the clever reader must now be asking the question `why did we start 
off using 3 different 
values of $\sqrt s$ -- why not 2 or 5?' This is a very important issue which we 
can only begin to address here. Let us return to the mass and couplings of 
case I and generate Monte Carlo `data' for only two values of 
$\sqrt s$=0.5 and 1 TeV with luminosities of $\cal L$= 100 and 220 $fb^{-1}$, 
respectively, thus keeping the total $\cal L$ the {\it same} as in the 
discussion above. Repeating our analysis we then 
arrive at the `2-point' fit as shown in Fig. 4a; unlike Fig. 1a, the 
allowed region in the leptonic coupling plane does not 
close and extends outward to ever larger values of 
$v_\ell,a_\ell$; we find that 
a similar result occurs for the $b$-quark couplings which are 
even more poorly determined. The corresponding 
$Z'$ mass contour is also found not to close, again extending outwards 
to ever larger $M_{Z'}$ 
values. We realize immediately that this is just what happens when data at 
only a single $\sqrt s$ is available. For our fixed $\cal L$, distributed as we 
have now done, we see that there is not a sufficient lever arm to 
simultaneously disentangle the $Z'$ mass and 
couplings. Of course the reverse situation can also be just as bad. We   
now generate Monte Carlo `data' for the case I mass and couplings in 100 GeV 
steps in $\sqrt s$ over the 0.5 to 1 TeV interval with the same total 
$\cal L$ as above but now distributed as 30, 30, 50, 50, 60, and 100 $fb^{-1}$, 
respectively. We then arrive at the `6-point' fit shown in Fig. 4b 
which suffers 
a problem similar to that presented in 
Fig. 4a. What has happened now is that we have spread 
the fixed $\cal L$ too thinly over too many points for the 
analysis to work. These same results are found to hold for all three cases. 
This brief study indicates that a proper balance is 
required to simultaneously achieve the desired statistics as well as an 
effective lever arm to obtain the $Z'$ mass and couplings. It is important 
to remember that we 
have {\it not} demonstrated that the `2-point' fit will never work. We note 
only that it fails with our specific fixed luminosity distribution for the 
masses and couplings associated with cases I-III. It is possible that for 
`lucky' combinations of masses and couplings a 2-point fit will suffice or it 
may work if substantially more luminosity is achievable. It is certainly true 
that all cases where at least three values of $\sqrt s$ are used 
will allow simultaneous mass and coupling extraction provided the 
integrated luminosity is available. 
Clearly, more work is required to further address these issues.  

\vspace*{-0.5cm}
\nn
\begin{figure}[htbp]
\centerline{
\psfig{figure=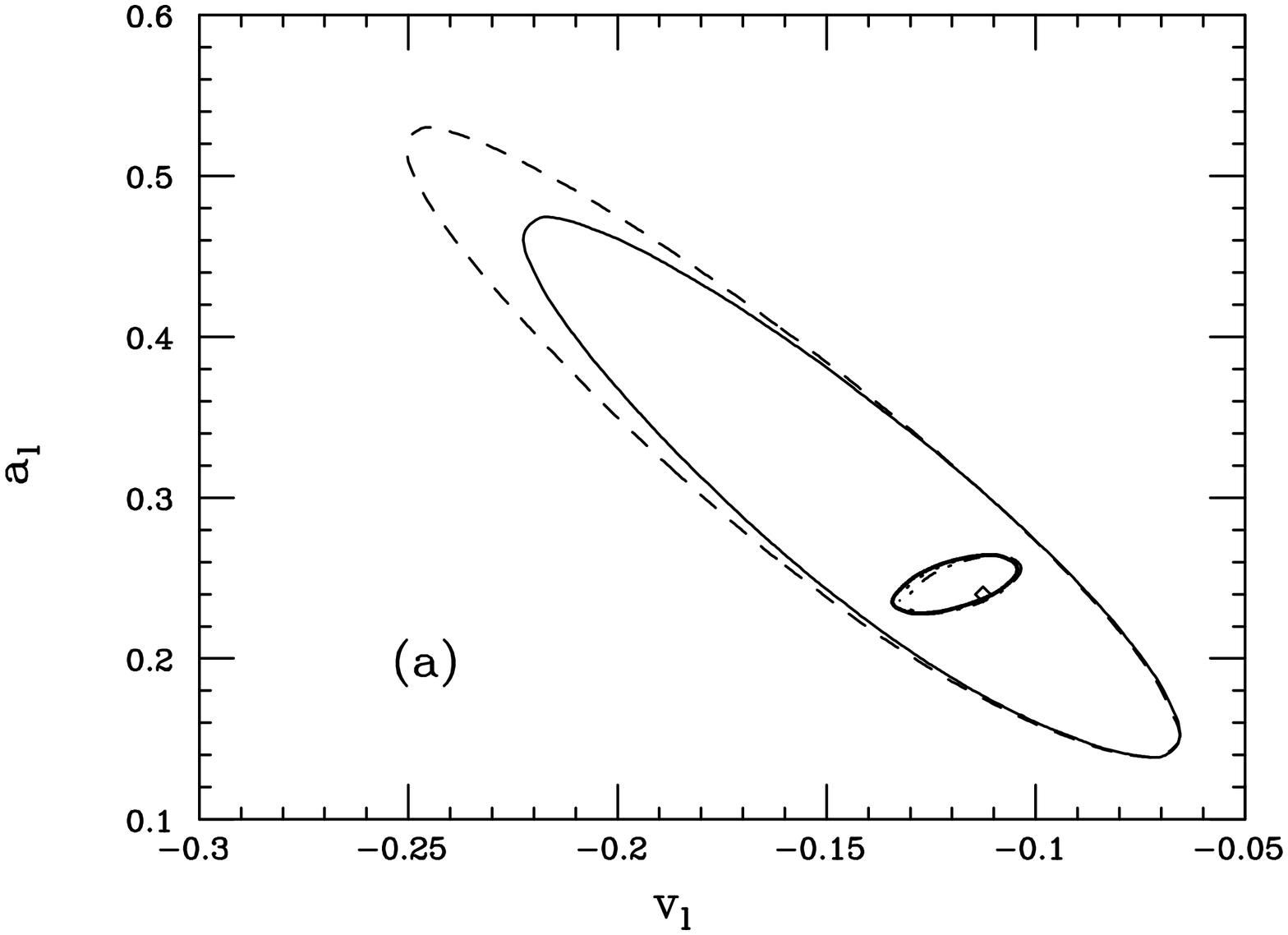,height=9.1cm,width=9.1cm,angle=-90}
\hspace*{-5mm}
\psfig{figure=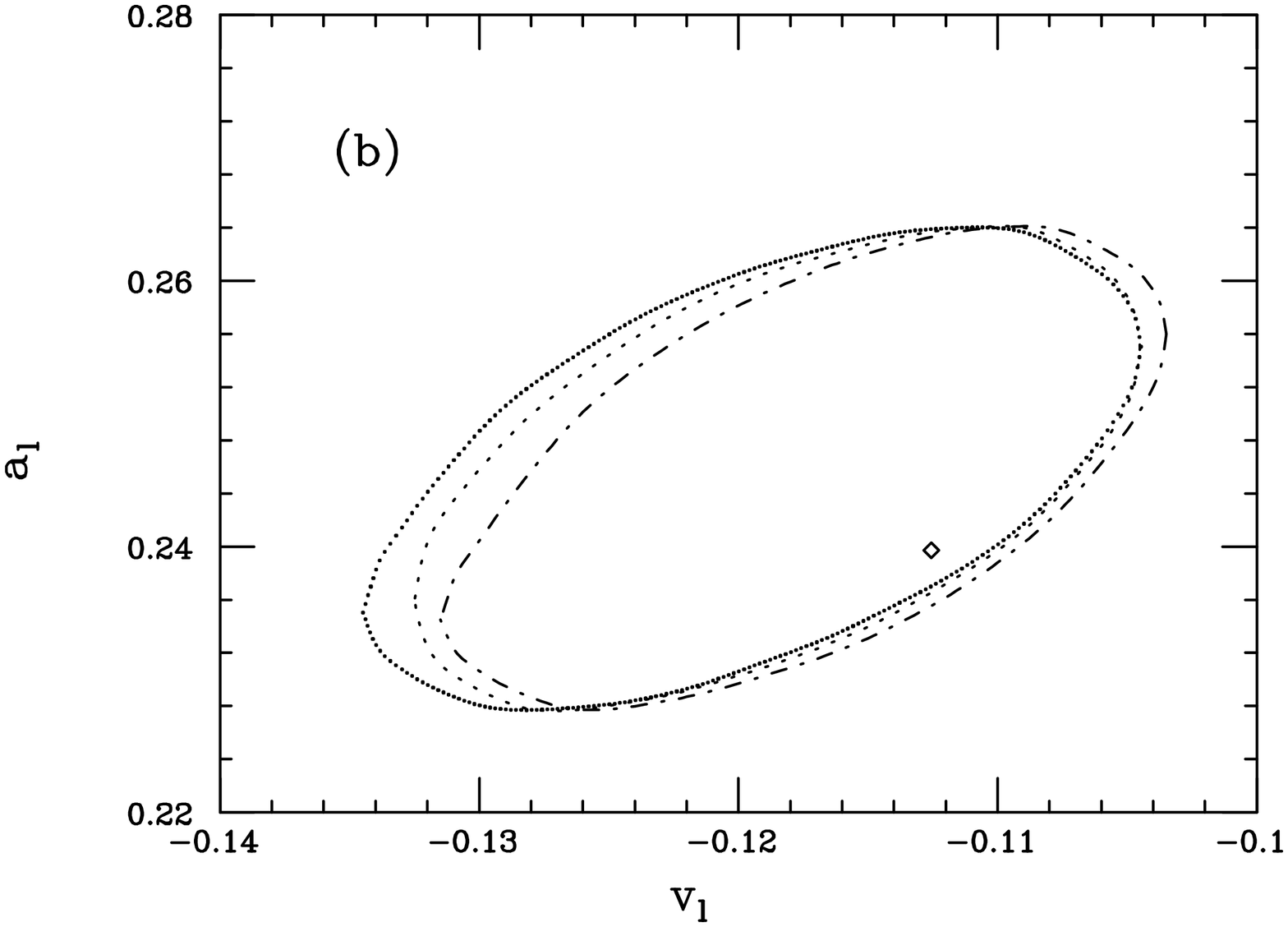,height=9.1cm,width=9.1cm,angle=-90}}
\vspace*{-0.75cm}
\centerline{
\psfig{figure=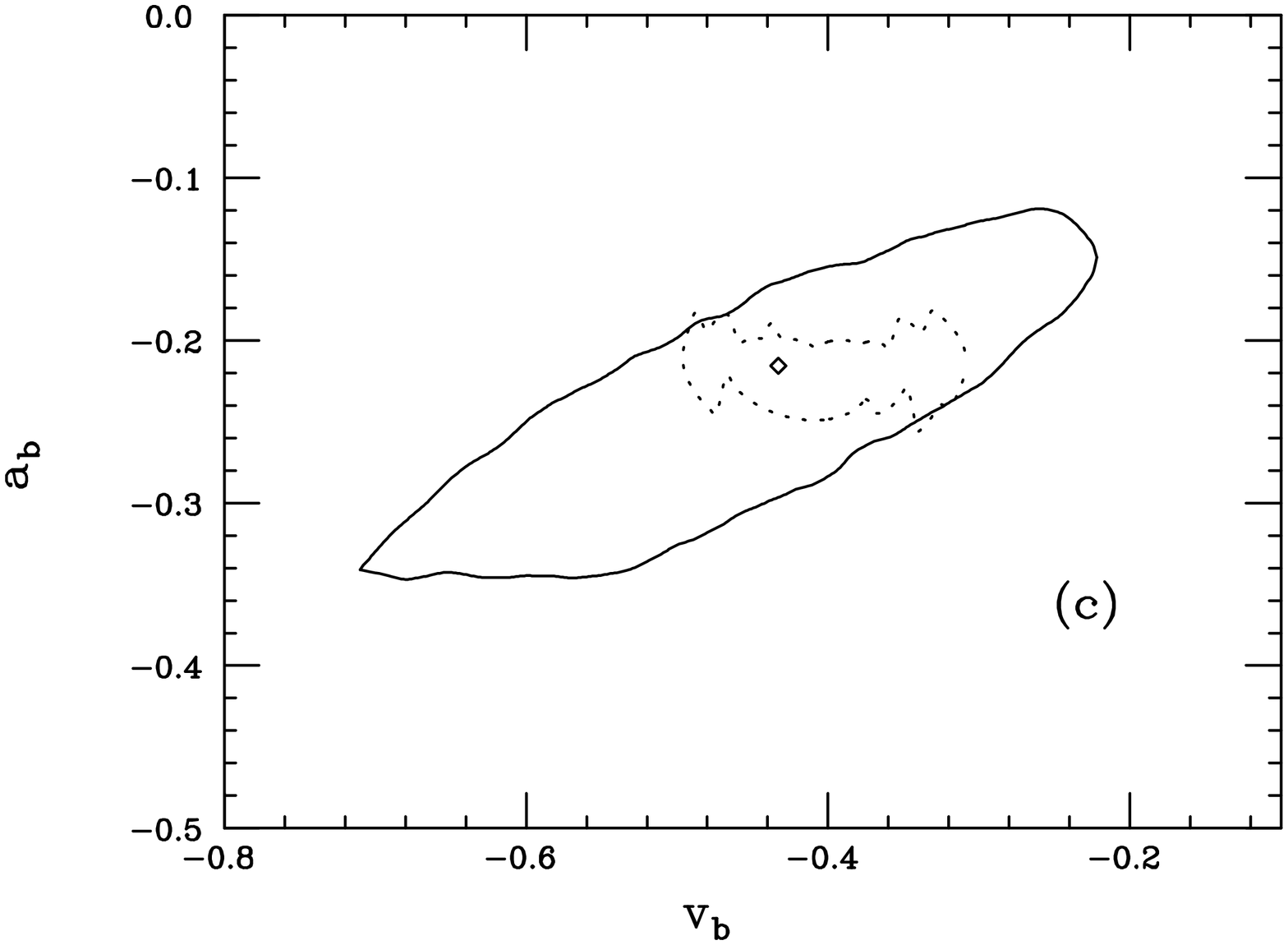,height=9.1cm,width=9.1cm,angle=-90}}
\vspace*{-1cm}
\caption{\small (a) Expanded lobe(solid) from Fig. 1a; the dashed curve shows 
the same result but for $P=80\%$. The smaller ovals, expanded in (b) apply 
when the $Z'$ mass is known. Here, in (b), $P=90(80)\%$ corresponds to the 
dash-dot(dotted) curve while the case of $P=90\%$ with $\delta P/P=5\%$ 
corresponds to the square-dotted curve. (c) Expanded lobe(solid) from Fig.1b; 
the dotted curve corresponds to the case when $M_{Z'}$ is known.} 
\end{figure}
\vspace*{-0.5cm}
\nn
\begin{figure}[htbp]
\centerline{
\psfig{figure=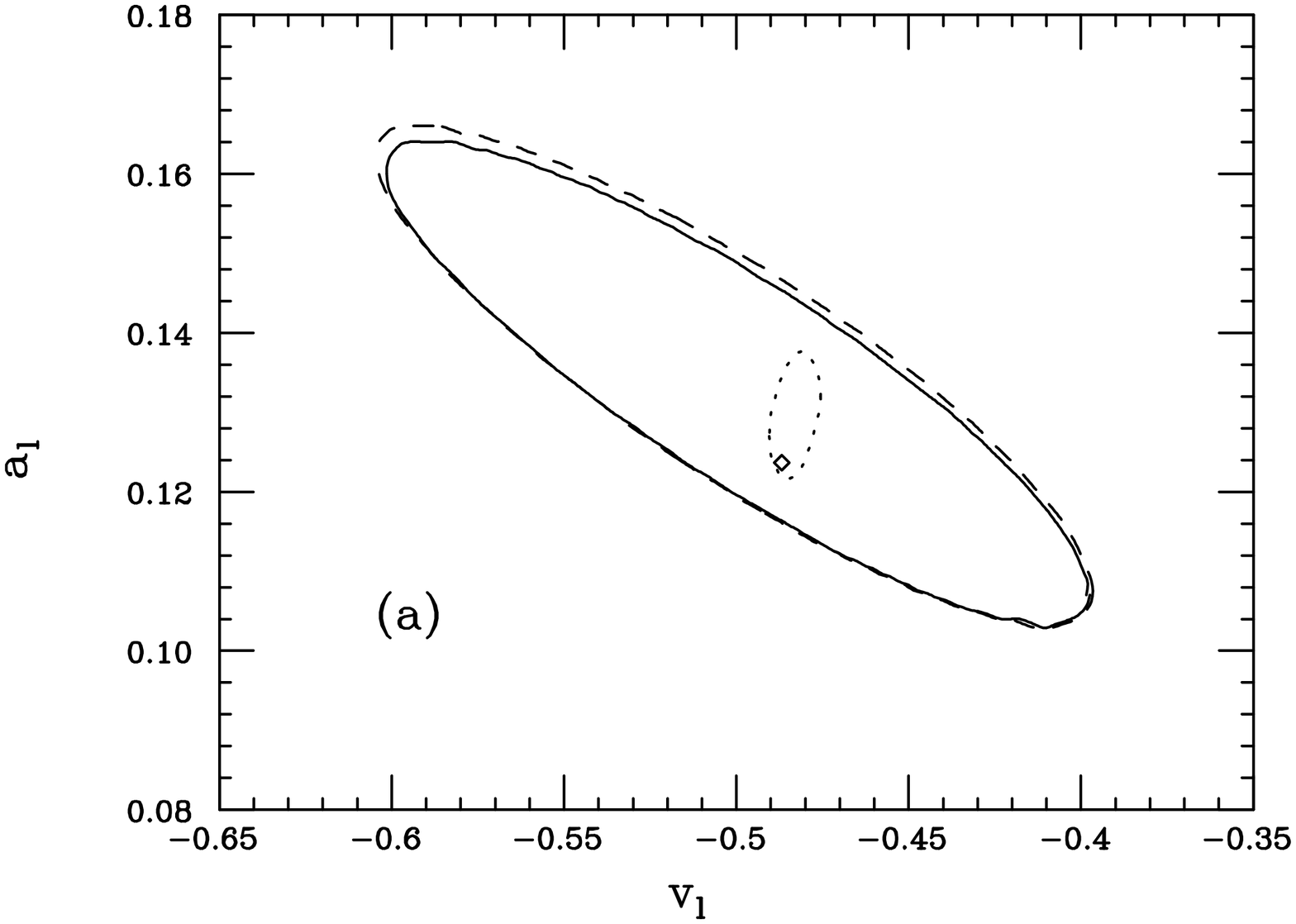,height=9.1cm,width=9.1cm,angle=-90}
\hspace*{-5mm}
\psfig{figure=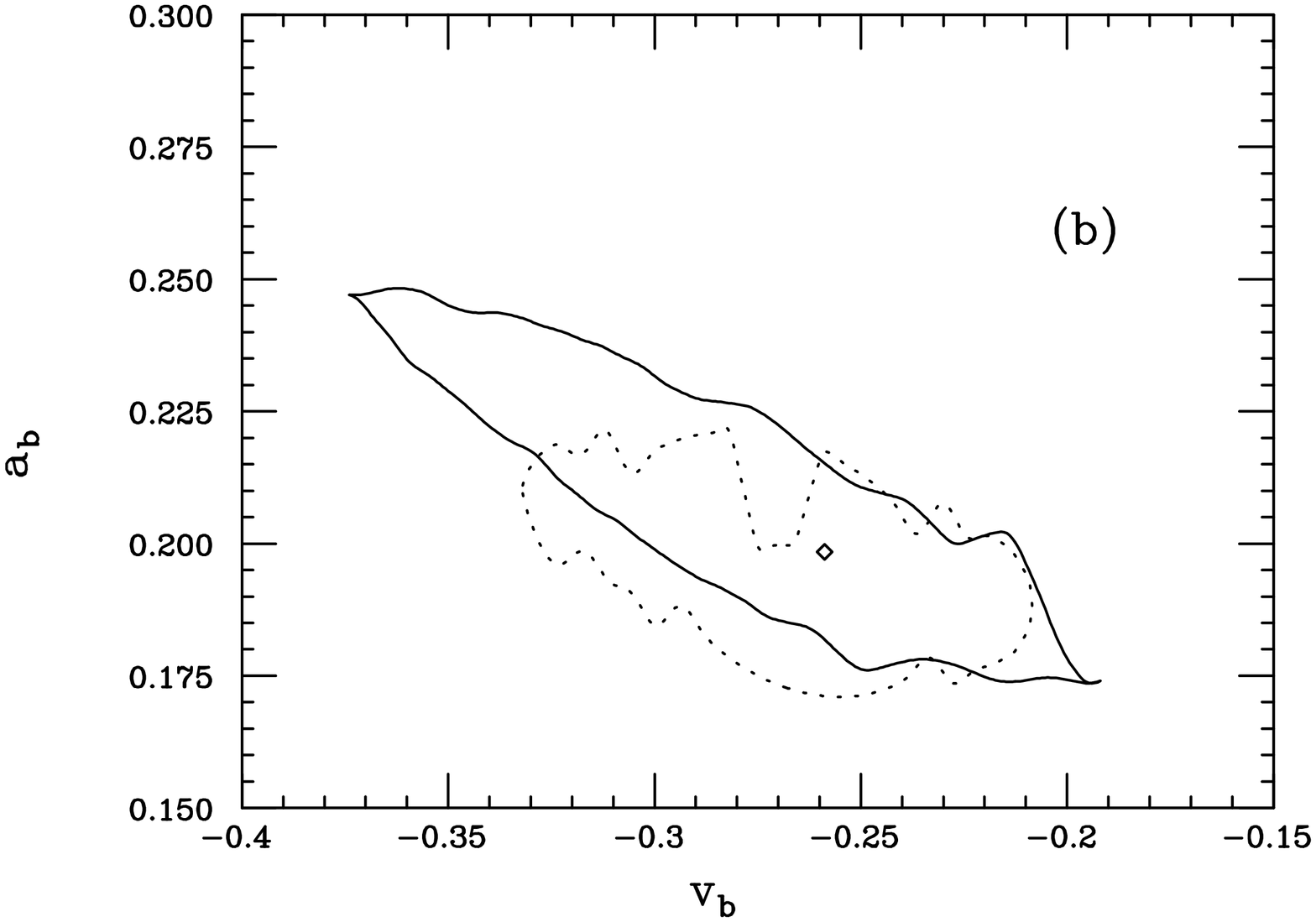,height=9.1cm,width=9.1cm,angle=-90}}
\vspace*{0.1cm}
\centerline{
\psfig{figure=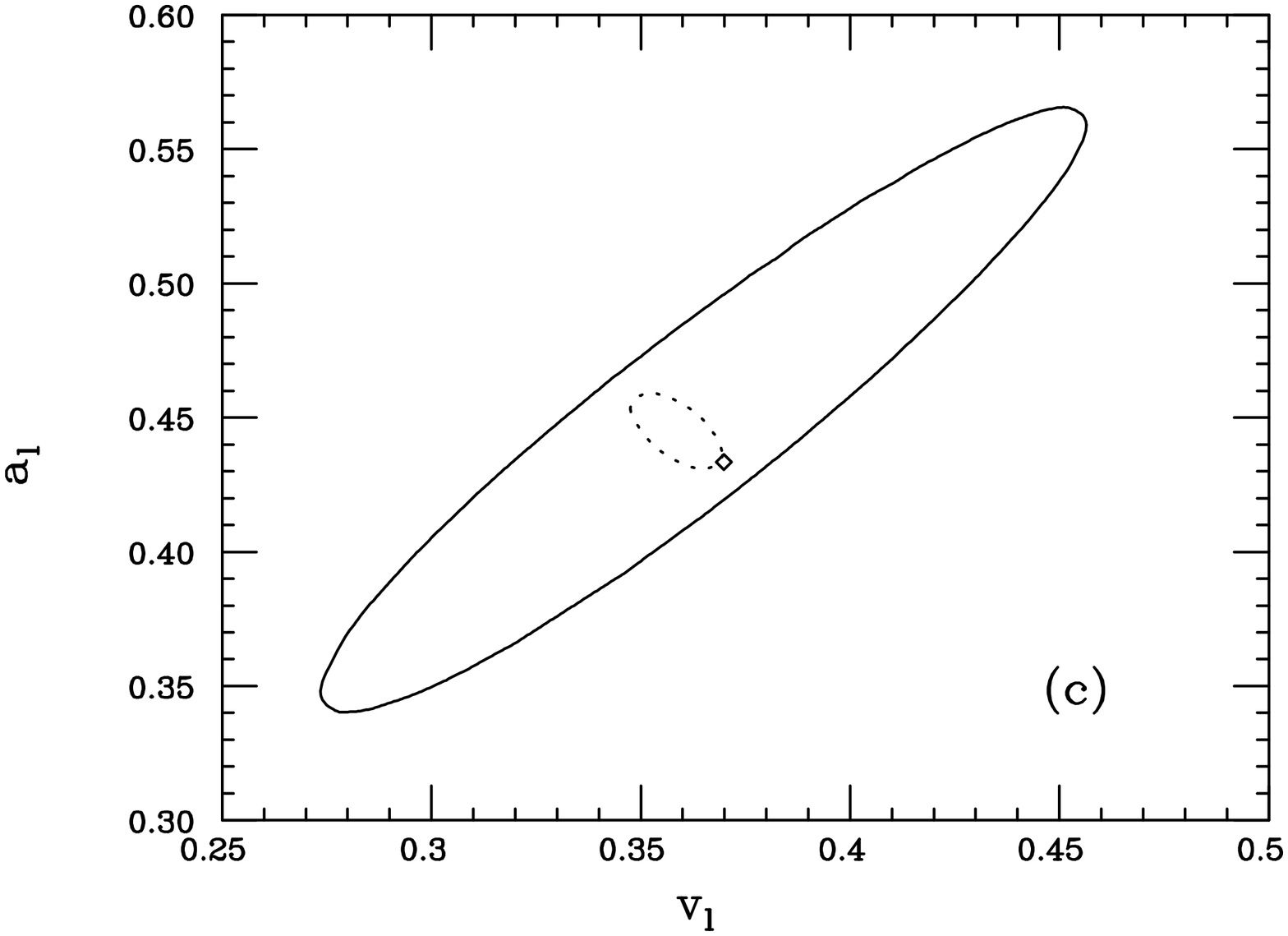,height=9.1cm,width=9.1cm,angle=-90}
\hspace*{-5mm}
\psfig{figure=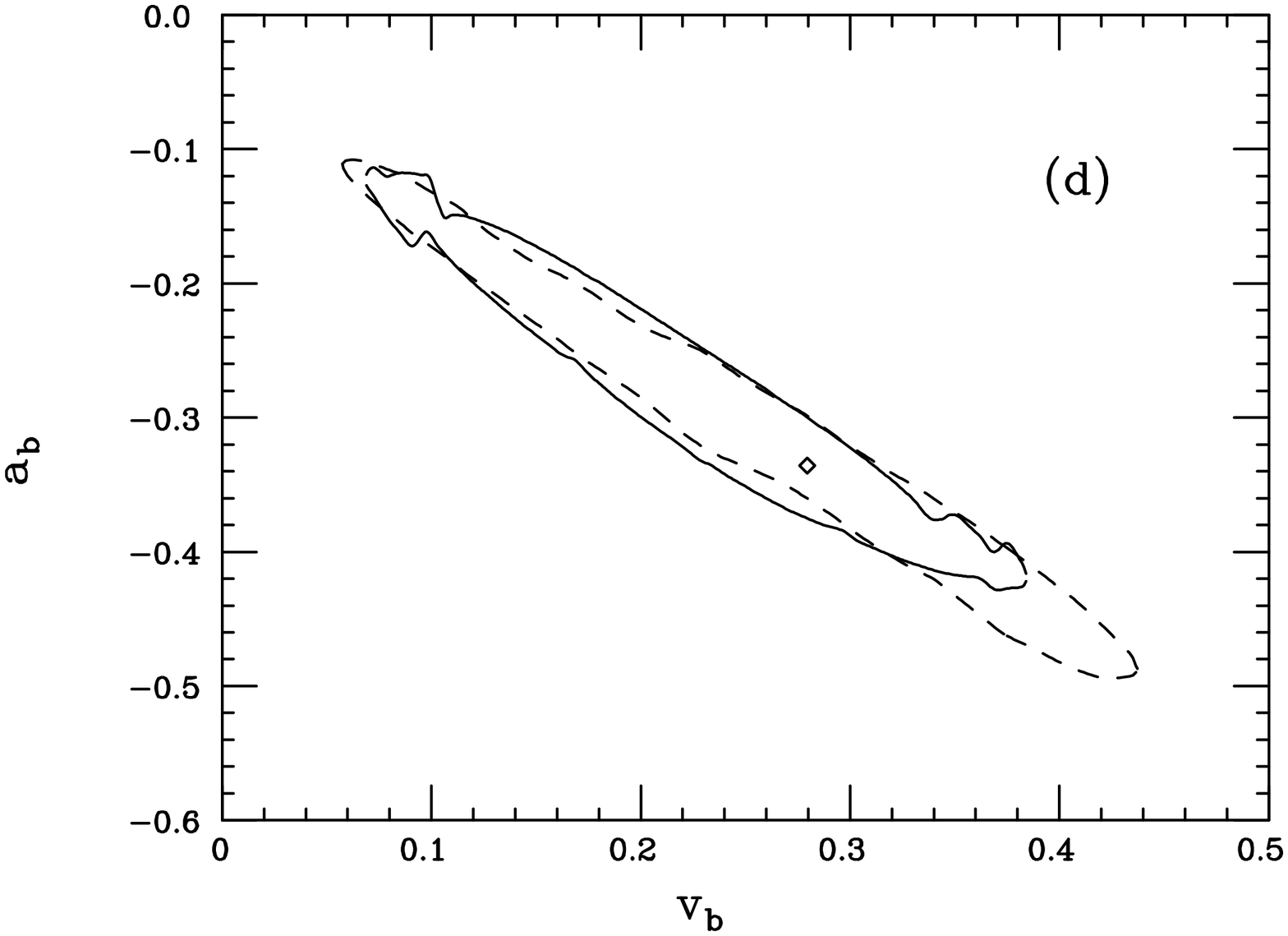,height=9.1cm,width=9.1cm,angle=-90}}
\vspace*{-1cm}
\caption{\small (a)Expanded lobe(solid) from Fig. 2a; the dashed curve shows 
the same result but for $P=80\%$. The smaller dotted oval, applies  
when the $Z'$ mass is known and $P=90\%$. (b) Expanded lobe(solid) from 
Fig. 2b; 
the dotted curve corresponds to the case when $M_{Z'}$ is known. (c) and (d) 
show the corresponding results for case III from Figs. 3a and 3b.}
\end{figure}
\vspace*{0.4mm}

How do these results change if $M_{Z'}$ {\it were} known or if our input 
assumptions were modified? In this case we use as additional input in our 
analysis the value of $M_{Z'}$ chosen by the Monte Carlo and perform 
four-dimensional fits to the same set of `data'. 
Let us return to case I and concentrate on the allowed 
coupling regions corresponding to a choice of negative values of 
$v_{\ell,b}$; these are 
expanded to the solid curves shown in Figs. 5a and 5c. (There will also be 
a corresponding region where $v_{\ell,b}$ are positive, which we ignore for 
the moment.) The large dashed curve 
in Fig. 5a corresponds to a reduction of the polarization to $80\%$ with the 
same relative error as before. While the allowed region expands the 
degradation is not severe. If the $Z'$ mass were known, the `large'  
ellipses shrink to 
the small ovals in Fig. 5a; these are expanded in Fig. 5b. This is clearly a 
radical reduction in the size of the allowed region. We see that when the 
mass is known, varying the polarization or its uncertainty over a reasonable 
range has very little influence on the resulting size of the allowed  
regions. From Fig. 5c we see that while knowing the $Z'$ mass significantly 
reduces the size of the allowed region for the $b$ couplings, the impact is 
far less than in the leptonic case 
for the reasons discussed above. Figs. 6a-d show that case I is not 
special in that similar results are seen to hold as well for cases II and III. 
For both of these cases, as in case I, there is an enormous reduction in the 
size of the allowed region for the leptonic couplings of the $Z'$ but the 
corresponding allowed region for the $b$-quark shrinks by about only a factor 
of two. In case III, there is hardly {\it any} reduction in the size of the 
allowed $b$-quark coupling region. 

\vspace*{-0.5cm}
\nn
\begin{figure}[htbp]
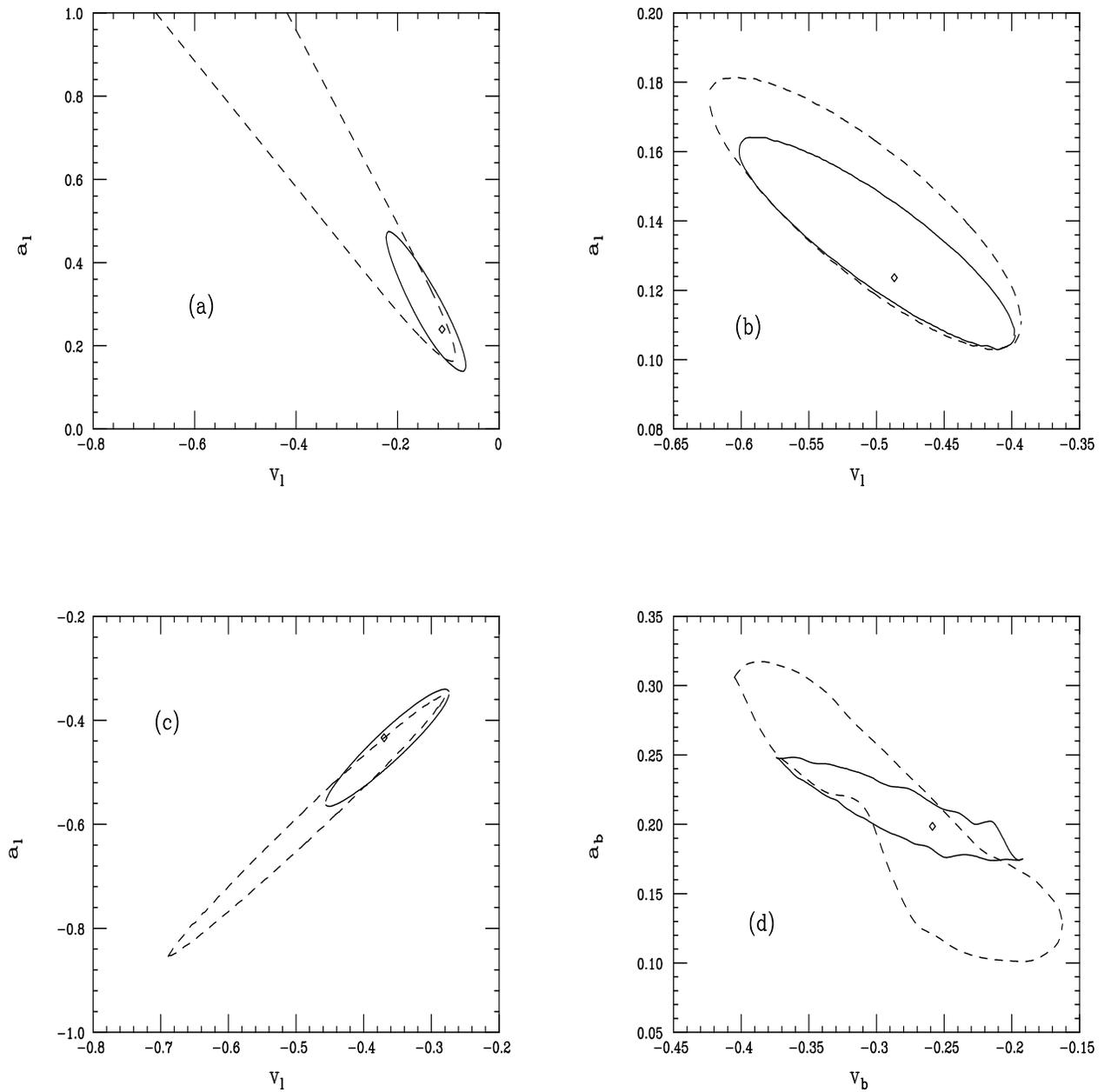

\centerline{
\psfig{figure=nlcnmuon.res1ps,height=9.1cm,width=9.1cm,angle=-90}
\hspace*{-5mm}
\psfig{figure=nlcnmuon.res2ps,height=9.1cm,width=9.1cm,angle=-90}}
\vspace*{0.1cm}
\centerline{
\psfig{figure=nlcnmuon.res3ps,height=9.1cm,width=9.1cm,angle=-90}
\hspace*{-5mm}
\psfig{figure=nlcnmuon.res4ps,height=9.1cm,width=9.1cm,angle=-90}}
\vspace*{-1cm}
\caption{A comparison of the constraints obtained on the $Z'$ leptonic 
couplings in cases I-III, in (a)-(c) respectively, both with(solid) and 
without(dashed) observables associated with beam polarization. In (d) are 
shown the corresponding $b$-quark couplings obtained for case II.} 
\end{figure}
\vspace*{0.4mm}

Just how large a role does large beam polarization play in obtaining our 
results? This becomes a critical issue, especially at higher energies, if 
our lepton collider is actually a muon collider where we may need to trade off 
luminosity for high beam polarization{\cite {nmc}}. As an extreme case, 
we repeat 
the previous analysis of cases I-III without including the observables 
associated with beam polarization in the fits; luminosities \etc ~remain the 
same as before. The results of this approach 
are shown in Figs. 7a-d. In case I, shown in Fig. 7a, the loss of the 
polarization dependent observables causes the allowed region in the leptonic 
coupling plane to open up and no closed region is found. Correspondingly, the 
$b$-quark couplings to the $Z'$ as well as the $Z'$ mass are constrained but 
are no longer localized. Somewhat better results are obtained in cases II and 
III, shown in Figs. 7b and 7c. For case II, the size of the allowed region 
essentially doubles in the $v_l-a_l$ plane and triples in the $v_b-a_b$ 
plane, as shown in Fig. 7d. The $M_Z'$ constraints are found to relax in a 
corresponding manner. In case II, while we are hurt by the lack of polarized 
beams we are still able to carry out the basic program of coupling extraction 
and $Z'$ mass determination -- unlike case I. In case III, Fig. 7c shows that the 
leptonic couplings are reasonably constrained without beam polarization but 
now neither the $Z'$ mass nor the $b$-quark couplings were found to 
be constrained. From these considerations we may conclude that 
beam polarization is of critical importance to this analysis unless the 
couplings lie in a `lucky' range. It is clear that for {\it arbitrary} 
values, we will not be able to simultaneously obtain mass and coupling 
determinations without large beam polarization. We note, however, that this 
conclusion can soften dramatically if the $Z'$ mass is already known.

What happens for larger $Z'$ masses or when data at larger values of $\sqrt s$ 
becomes available? (As stated above, the `reach' in our coupling 
determinations was $\simeq 2$ TeV using the `data' at 500, 750 and 1000 GeV.) 
Let us assume that the `data' from the above three center 
of mass 
energies is already existent, with the luminosities as given. We now imagine 
that the NLC increases its center of mass energy to $\sqrt s$= 1.5 TeV and 
collects an additional 200 $fb^{-1}$ of integrated luminosity, which 
corresponds to 1-2 design years. 
Clearly for $Z'$ masses near or below 1.5 TeV our 
problems are solved since an on-shell $Z'$ can now be produced. Thus we 
shall concern ourselves only with $Z'$ masses in excess of 2 TeV, inaccessible 
in the lower energy study above. 
Figs. 8a-d  show the result of extending our previous procedure--now using 4 
different $\sqrt s$ values, for two distinct choices (IV and V) of the 
$Z'$ mass and 
couplings. These `4-point' results are a combined fit to the data at 
all four center of mass energies. We show the results for both the general 
case where $M_{Z'}$ is unconstrained as well as when it is already 
determined by other data. 
(As before, only one of the allowed pair of ellipses resulting from the 
overall sign ambiguity is shown for simplicity.) 
Note that the $Z'$ input masses we have chosen are well in excess of 2 TeV 
where the LHC may provide only very minimal information on the fermion 
couplings{\cite {rev1}}. Clearly by using the additional data from a 
run at $\sqrt s$=1.5 TeV this technique can be extended to perform coupling 
extraction for $Z'$ masses in excess of 2.5 TeV. The maximum `reach' for the 
type of coupling analysis we have performed is not yet determined. It seems 
likely, based on these initial studies, that the extraction of interesting 
coupling information for $Z'$ masses in excess of 3 TeV may be possible for a 
reasonable range of coupling parameters. 

\vspace*{-0.5cm}
\nn
\begin{figure}[htbp]
\centerline{
\psfig{figure=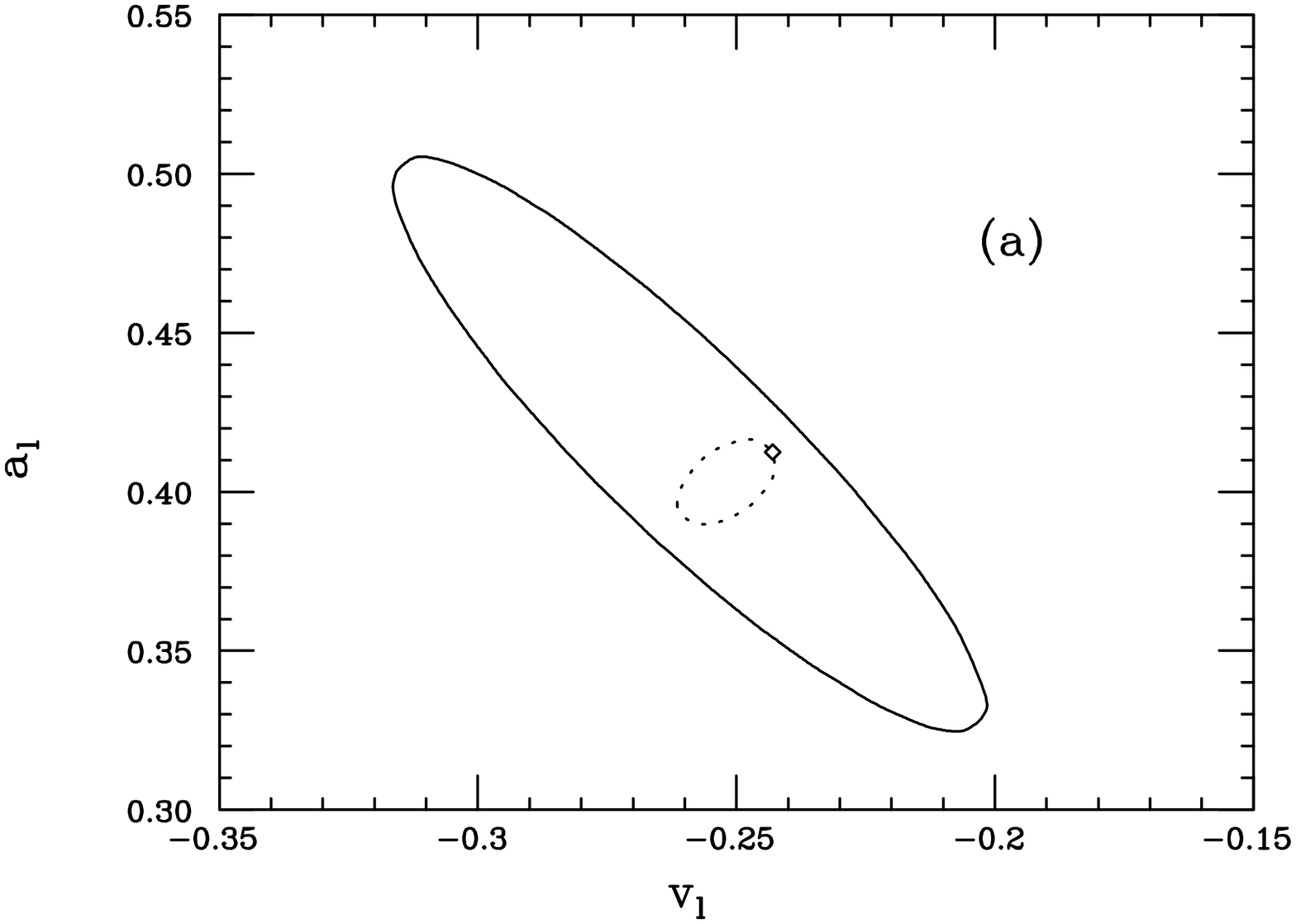,height=9.1cm,width=9.1cm,angle=-90}
\hspace*{-5mm}
\psfig{figure=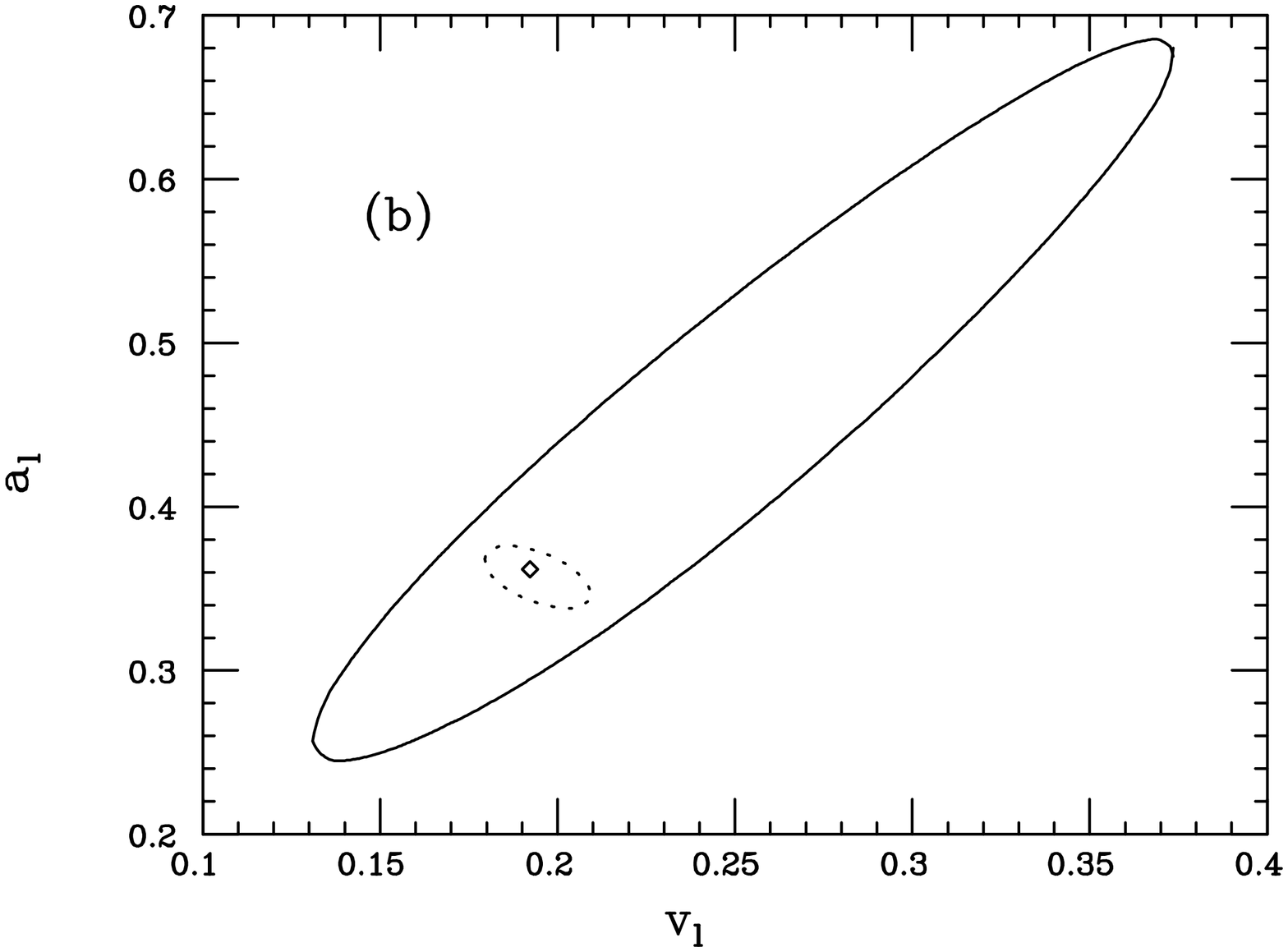,height=9.1cm,width=9.1cm,angle=-90}}
\vspace*{0.1cm}
\centerline{
\psfig{figure=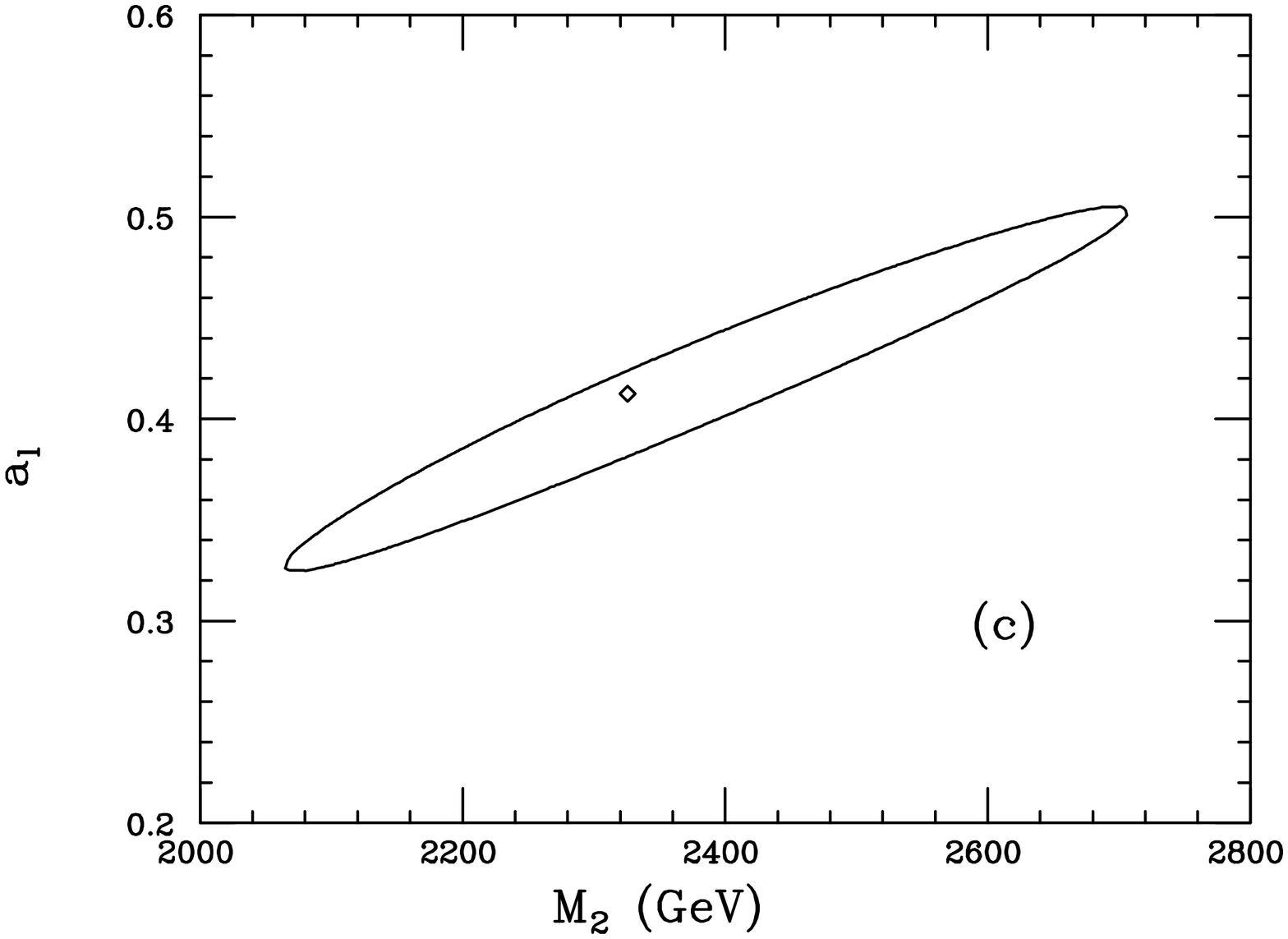,height=9.1cm,width=9.1cm,angle=-90}
\hspace*{-5mm}
\psfig{figure=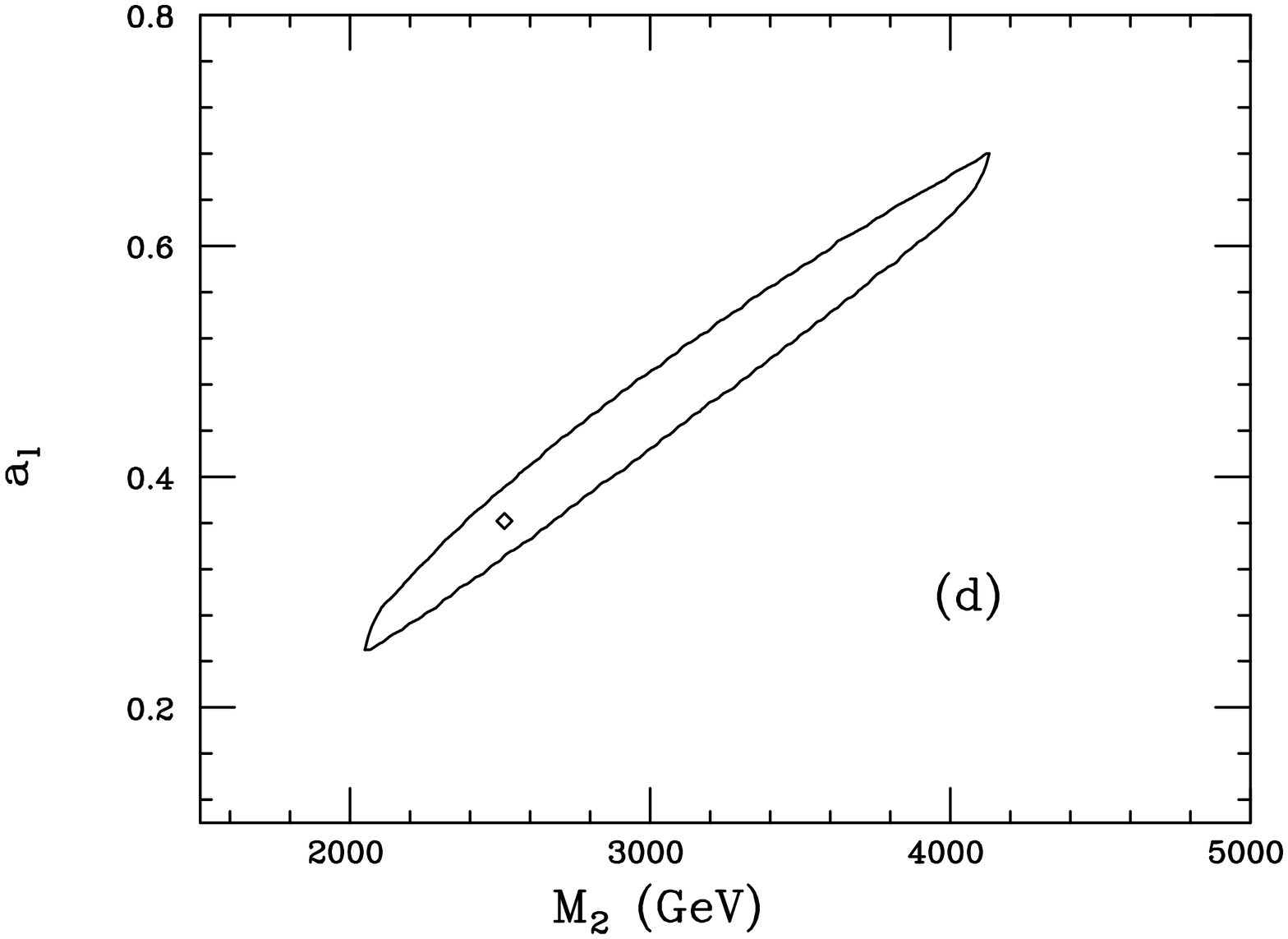,height=9.1cm,width=9.1cm,angle=-90}}
\vspace*{-1cm}
\caption{\small Lepton 
coupling determination for $Z'$'s with masses of (a) 2.33 TeV and (b) 2.51 TeV 
when the mass is unknown(solid) and known(dotted) corresponding to cases IV 
and V discussed in the text. (c) and (d) are the 
corresponding mass determinations which result from the five-dimensional 
fit. These results include an 
additional 200 $fb^{-1}$ of luminosity taken at a center of mass energy of 
1.5 TeV.} 
\end{figure}
\vspace*{0.4mm}

\section{{\bf Outlook and Conclusions}}

In this paper we have shown that it is possible for the NLC to extract 
information on the $Z'$ couplings to leptons and $b$-quarks even when the $Z'$ 
mass is not {\it a priori} known and, in fact, determine the $Z'$ mass. This 
has been demonstrated in a model-independent manner by randomly and 
anonymously choosing the mass and couplings of the $Z'$ and demonstrated the 
power of precision measurements at future linear colliders. 
The critical step for the success of the 
analysis is to combine the data available from measurements performed 
at several different center of mass energies. For a reasonable distribution of 
the luminosities 
the specific results we have obtained suggest, but do not prove, that data 
sets obtained at at least 3 different energies are necessary for the 
procedure to be successful. The mass `reach' for this approach is 
approximately twice the highest center of mass energy available. Several 
question remain about the optimization of our approach; these will be 
addressed in future work.

\vskip.25in
\section{{\bf Acknowledgements}}

The author would like to thank J.L. Hewett, S. Godfrey, S. Riemann,  
K. Maeshima and H. Kagan for discussions related to this work.

\newpage

%
\def\MPL #1 #2 #3 {Mod.~Phys.~Lett.~{\bf#1},\ #2 (#3)}
\def\NPB #1 #2 #3 {Nucl.~Phys.~{\bf#1},\ #2 (#3)}
\def\PLB #1 #2 #3 {Phys.~Lett.~{\bf#1},\ #2 (#3)}
\def\PR #1 #2 #3 {Phys.~Rep.~{\bf#1},\ #2 (#3)}
\def\PRD #1 #2 #3 {Phys.~Rev.~{\bf#1},\ #2 (#3)}
\def\PRL #1 #2 #3 {Phys.~Rev.~Lett.~{\bf#1},\ #2 (#3)}
\def\RMP #1 #2 #3 {Rev.~Mod.~Phys.~{\bf#1},\ #2 (#3)}
\def\ZP #1 #2 #3 {Z.~Phys.~{\bf#1},\ #2 (#3)}
\def\IJMP #1 #2 #3 {Int.~J.~Mod.~Phys.~{\bf#1},\ #2 (#3)}

\end{document}